\documentclass{aa}

\usepackage{graphicx}
\usepackage{txfonts}
\usepackage{makecell}
\usepackage{pifont}
\usepackage{bm}
\usepackage[fleqn]{amsmath}
\usepackage{subcaption}
\usepackage{lscape}
\usepackage{placeins}
\usepackage{multirow}
\usepackage{hyperref}
\newcommand{\cmark}{\ding{51}}
\newcommand{\xmark}{\ding{55}}

\begin{document}

\allowdisplaybreaks

   \title{Coverage is not enough: Frequentist tests of simulation-based inference for primordial non-Gaussianity}

   \author{Toka Alokda\inst{1,2}\thanks{\email{talokda@astro.uni-bonn.de}}
        \and Cristiano Porciani\inst{1,2} \and Alexander Eggemeier\inst{1,2}}

   \institute{Argelander-Institut für Astronomie, Universität Bonn, Auf dem Hügel 71, 53121 Bonn, Germany \and Cluster of Excellence ``Our Dynamic Universe'' (DYNAVERSE), Institute of Physics I, University of Cologne, Zülpicher Str. 77, 50937 Cologne, Germany}

   \date{Received 28 April 2026 / Accepted 11 August 2026}

  \abstract
{Simulation-based inference (SBI) has emerged as a powerful framework for extracting cosmological information from complex, nonlinear data where analytical likelihoods are unavailable. Its reliability is commonly assessed using coverage-based diagnostics, which test whether posterior credible regions contain the true parameter with the expected frequency under the prior predictive distribution. However, these tests probe calibration only in an averaged sense and do not constrain posterior behavior at a fixed parameter value, which is the regime relevant for practical inference where the data are generated at a
single (unknown) truth.}
{We investigated these limitations in the context of constraining primordial non-Gaussianity, parameterized by $f_\mathrm{NL}$, using
simulations of the dark matter halo field. In particular, we assessed whether calibration under standard diagnostics ensures reliable uncertainty quantification at a fixed parameter value.}
{We compared SBI based on contrastive neural ratio estimation (CNRE) with likelihood-based inference (LBI) using three complementary summary statistics: the power spectrum, bispectrum, and wavelet scattering transform (WST) coefficients. Posterior distributions were evaluated across 1000 independent test simulations.}
{SBI and LBI agree well on posterior means and skewness, while the variance agrees on average but it shows a weaker realization-by-realization consistency. Larger differences arise in the kurtosis, indicating discrepancies in the posterior tails. These effects are already present for the power spectrum---where the Gaussian likelihood assumed in our LBI analysis is best justified---and they are most pronounced for the combined power spectrum and bispectrum, where SBI posteriors are often underconfident and can yield weaker constraints than either statistic individually, despite passing coverage tests. WST coefficients further tighten constraints on $f_\mathrm{NL}$ even when restricting the analysis to large scales ($k_\mathrm{max}=0.141\,h\,\mathrm{Mpc}^{-1}$).}
{Our results highlight both the potential of higher-order statistics for probing primordial non-Gaussianity and the limitations of current validation strategies for SBI. Diagnostics sensitive to posterior shape and behavior at a fixed parameter value will be necessary for robust uncertainty quantification in precision cosmology.}


   \maketitle
   \nolinenumbers

\section{Introduction}

Simulation-based inference (SBI) has become an increasingly important tool for extracting cosmological information from complex and nonlinear datasets. By replacing an explicit likelihood with a neural approximation to the likelihood, likelihood–evidence ratio, or posterior distribution, SBI enables parameter inference even when analytical likelihoods are unavailable or impractical to compute \citep{Kcranmer_sbi,deistler2025simulationbasedinferencepracticalguide}. 
This capability makes SBI particularly attractive for modern large-scale structure (LSS) analyses, where the observables are highly nonlinear and their statistical properties are only partially understood.

Despite these advantages, ensuring the reliability of SBI posterior estimates remains a major challenge. Neural posterior estimators can produce miscalibrated or overconfident posteriors when the training data fail to sufficiently represent the space of possible observations, or when the chosen model lacks expressive power \citep{Hermans2021averting, pmlr-v130-lueckmann21a}. Several strategies have been developed to mitigate these effects, including classifier-based likelihood-ratio estimation \citep{CNRE_papera, cnre_paperb}, explicit calibration diagnostics based on coverage tests \citep{talts_SBC, TARP_test}, and training schemes designed to penalize overconfidence \citep{Hermans2021averting, bnre_paper}. 

Coverage-based diagnostics have become the standard tool for validating SBI pipelines. These tests assess whether posterior credible regions contain the true parameter value with the expected frequency under repeated simulations drawn from the prior predictive distribution. However, they probe calibration only in an averaged sense and do not constrain the behavior of the posterior at a fixed parameter value. As a result, an inference method can satisfy such diagnostics while still exhibiting systematic biases or the misestimation of uncertainties in the regime relevant for practical applications.

To investigate this issue in a controlled setting, we considered the inference of primordial non-Gaussianity (PNG), parameterized by $f_\mathrm{NL}$. PNG describes deviations from Gaussian initial conditions and provides a sensitive probe of early-Universe physics \citep{Gangui_1994}. While the cosmic microwave background currently provides the tightest constraints \citep{PLANCK2018}, LSS offers a complementary probe through the scale-dependent clustering it induces in dark matter halos and galaxies \citep{Dalal_2008, Slosar_2008}. This signal can be extracted using summary statistics such as the power spectrum and bispectrum, as well as more recent nonlinear descriptors such as the wavelet scattering transform  \citep[WST,][]{wst_network, Valogiannis_2022b,Valogiannis_2022a,Lemos_2024,Peron_2024}. Because the likelihoods of these summaries are generally unknown, they provide a natural testbed for SBI.

For this study, we used simulations of the dark matter halo density field to investigate how SBI performs in the context of estimating $f_\mathrm{NL}$ from different LSS observables. We focused on three complementary summary statistics: the power spectrum, bispectrum, and WST coefficients. We compared posterior estimates obtained with SBI using contrastive neural ratio estimation (CNRE) with those obtained from traditional likelihood-based inference (LBI). Using standard calibration and validation procedures, we assessed the reliability of current SBI pipelines when confronted with realistic cosmological data models.

The structure of this paper is as follows. Section~\ref{sec:inference} provides an overview of the inference frameworks and validation strategies adopted in this work. Section~\ref{sec:methods} describes the simulations, the computation of summary statistics, and the implementation of both the LBI and SBI pipelines. The results are presented and discussed in Section~\ref{sec:results}, followed by the conclusions in Section~\ref{sec:conclusions}.

\section{Simulation-based inference}\label{sec:inference}

Bayesian inference provides a framework to estimate cosmological parameters by updating prior knowledge with observational data. According to Bayes' theorem, the posterior distribution of parameters $\theta$ given data $x$ is
\begin{align}
p(\theta\mid x) = \frac{p(x\mid\theta)\, p(\theta)}{p(x)},
\label{eq:bayes}
\end{align}
where $p(\theta)$ is the prior, $p(x\mid\theta)$ the likelihood, and $p(x)$ the evidence.

In many cosmological applications the likelihood $p(x\mid\theta)$ is unknown or computationally intractable, while simulations that generate data $x\sim p(x\mid\theta)$ remain available. SBI, also known as likelihood-free or implicit-likelihood inference, provides a framework for performing Bayesian inference in this setting by learning the relationship between simulated data and model parameters \citep{Kcranmer_sbi,deistler2025simulationbasedinferencepracticalguide}.

Neural simulation-based inference uses neural networks to approximate quantities appearing in Bayes' theorem using simulated training data $(\theta,x)\sim p(\theta)\,p(x\mid\theta)$. Depending on the quantity that is learned, three main approaches arise: neural posterior estimation (NPE), neural likelihood estimation (NLE), and neural ratio estimation (NRE), which we focus on here.

\subsection{Neural ratio estimation}
\label{subsec:nre}

Neural ratio estimation directly learns the likelihood-to-evidence ratio
\begin{align}
r(x,\theta)=\frac{p(x\mid\theta)}{p(x)},
\end{align}
from which the posterior follows as $p(\theta\mid x)\propto r(x,\theta)\,p(\theta)$.

In practice, the ratio is estimated by training a binary classifier $d_\phi(x,\theta)$, where $\phi$ denotes the parameters of the neural network. to distinguish between samples drawn from the joint distribution $p(\theta)p(x\mid\theta)$ and samples drawn from the product of marginals $p(\theta)p(x)$ \citep{CNRE_papera,cnre_paperb,bnre_paper}. The classifier is trained using the cross-entropy loss
\begin{align}
\mathcal{L}_{\mathrm{NRE}}(\phi)
=
-\mathbb{E}_{p_{\text{joint}}}[\log d_\phi(x,\theta)]
-\mathbb{E}_{p_{\text{indep}}}[\log(1-d_\phi(x,\theta))].
\end{align}

At the optimal solution, the classifier recovers the likelihood-to-evidence ratio through the log-odds relation
\begin{align}
\log r_{\phi^\star}(x,\theta)
=
\log\frac{d_{\phi^\star}(x,\theta)}{1-d_{\phi^\star}(x,\theta)} .
\end{align}

Different SBI methods have been benchmarked extensively
\citep{pmlr-v130-lueckmann21a}, showing that although neural
network-based approaches generally outperform classical techniques, no
single method consistently performs best across all inference problems. 
We adopt NRE over NLE and NPE because its classification-based objective avoids explicit density normalization, making it easier to optimize for high-dimensional summary statistics \citep{Kcranmer_sbi,deistler2025simulationbasedinferencepracticalguide}. We use the contrastive variant, CNRE, which extends the binary objective to multiple negative samples. This improves the stability and efficiency of ratio estimation, particularly in high-dimensional settings \citep{CNRE_papera,cnre_paperb}. 
Details of our implementation are given in Section~\ref{subsec:sbi_implement}.

Several open-source software packages implement state-of-the-art SBI algorithms, including the widely used \textsc{sbi} package
\citep{BoeltsDeistler_sbi_2025} and the \textsc{LtU-ILI} package
\citep{Ho_2024}, which is tailored to astrophysical and cosmological
applications. We use \textsc{sbi} because it provides a flexible
framework for comparing multiple inference algorithms and training
strategies within a common interface.

\subsection{Posterior calibration and coverage tests}

Assessing the reliability of posterior distributions inferred with simulation-based methods requires dedicated diagnostics. A central concept in this context is coverage.

A credible region $C_\alpha(x)$ with posterior probability mass (or credibility level) $\alpha$ is defined as any subset of parameter space satisfying\footnote{Credible regions are not unique. Common choices include central (equal-tailed) intervals defined by posterior quantiles and highest posterior density regions.}
\begin{equation}
\int_{C_\alpha(x)} p(\theta|x)\,\mathrm{d}\theta = \alpha.
\end{equation}

In practice, statistical models are only approximations of reality, and posterior statements should therefore be interpreted as conditional on the assumed model. To assess internal consistency, one can consider an idealized setting in which the model is correct. In this case, data can be generated from the prior predictive distribution
\begin{equation}
    p(x) = \int p(\theta,x)\,\mathrm{d}\theta=\int p(x|\theta)\,p(\theta)\,\mathrm{d}\theta
\end{equation}
which is obtained from the joint distribution of parameters and data implied by the model. Sampling from this distribution provides a controlled setting in which the inference procedure can be validated.

The coverage probability of $C_\alpha(x)$ is defined as the fraction of realizations for which the credible region contains the true parameter value when $(\theta,x)$ are drawn from the prior predictive distribution. A posterior is said to be "well calibrated" if, for any credibility level $\alpha$,
\begin{equation}
    \mathbb{P}\big[\theta \in C_\alpha(x)\big] = \alpha,
\end{equation}
where the probability is taken over $p(\theta,x)$. This property reflects the self-consistency of Bayesian inference under the prior predictive distribution and provides a natural frequency-based validation criterion.

A standard approach to assessing coverage is "simulation-based calibration" \citep[SBC,][]{talts_SBC}. In this framework, pairs $(\theta_i, x_i)$ are drawn from the prior and the forward model, the corresponding posteriors are computed, and the rank of the true parameter value is evaluated within samples drawn from each posterior. If the posterior is correctly calibrated, these ranks follow a uniform distribution. Equivalently, credible intervals contain the true parameter with the nominal frequency.

While SBC provides a robust test of marginal calibration, it can be insensitive to errors in the full structure of the posterior, particularly in high-dimensional settings. More stringent diagnostics, such as the Tests of Accuracy with Random Points \citep[TARP,][]{TARP_test}, have therefore been proposed. In this approach, calibration is assessed by evaluating coverage with respect to regions defined relative to randomly chosen reference points in parameter space.

Specifically, for a reference point $r$ and true parameter value $\theta_0$, define the region
\begin{equation}
    C_{r,\theta_0} = \{\theta : \|\theta - r\| \le \|\theta_0 - r\|\},
\end{equation}
where $\|\cdot\|$ denotes a distance metric in parameter space. The TARP condition requires that the expected coverage probability (ECP) matches the posterior mass of $ C_{r,\theta_0}$, i.e.,
\begin{equation}
\mathbb{P}\big[\theta_0 \in C_{r,\theta_0}(x)\big]
=
\int_{C_{r,\theta_0}} p(\theta|x)\,\mathrm{d}\theta,
\end{equation}
for all $r$. In contrast to standard coverage, which fixes the probability level $\alpha$ and varies the region, TARP fixes the region and compares its posterior mass to its empirical frequency.

By probing a broader class of regions, TARP is more sensitive to mis-specification of the posterior shape, including biases, incorrect dispersion, or distorted tails, that may not be detected by marginal coverage alone. In the ideal limit, satisfying the TARP conditions for all reference points provides a necessary and sufficient condition for posterior correctness, i.e., $q_{\phi^{\star}}(\theta|x) = p(\theta|x)$ almost everywhere with respect to the prior predictive distribution.

In practice, however, these tests are limited by finite sampling (finite simulations, posterior samples, and number of reference points) and incomplete coverage of the parameter space. As a result, passing them does not strictly guarantee full posterior accuracy in realistic settings.

\subsection{Frequentist testing}

In practical applications, assuming the model provides an adequate description of the data-generating process, observations can be viewed as arising from a single (unknown) set of true parameter values $\theta_0$. It is therefore important to assess the behavior of posterior estimates under repeated sampling at fixed truth, i.e., for realizations $x \sim p(x|\theta_0)$. Frequentist diagnostics, such as coverage at fixed parameter values, provide a direct probe of the reliability of uncertainty quantification in this setting.

By construction, SBC and TARP assess calibration under the prior predictive distribution, i.e., averaged over both parameters and data, and therefore do not directly constrain the behavior of posterior estimates at fixed parameter value. In particular, the subspace $\theta=\theta_0$ has zero measure under the joint distribution $p(\theta,x)$, so that properties that hold almost everywhere under the prior predictive distribution need not hold when conditioning on a fixed parameter value. As a result, posterior correctness in the prior-predictive sense does not, in general, guarantee correct behavior at fixed truth. Although it is sometimes argued that smoothness or asymptotic considerations should allow one to extend prior-predictive validity to fixed-parameter settings, this argument is not generally justified, and systematic deviations can arise in practice.

In the remainder of this work, we use such frequentist tests to compare LBI and SBI methods. We will show that, even when SBI satisfies standard calibration diagnostics such as SBC and TARP, its posterior estimates can differ substantially from those obtained with LBI when evaluated at fixed parameter value.

 \section{Primordial non-Gaussianity as a test case}\label{sec:methods}

As a test case for comparing likelihood-based and simulation-based inference, we consider constraints on PNG using LSS statistics. PNG encodes deviations from Gaussian statistics in the primordial curvature perturbations generated in the early Universe. In the simplest single-field slow-roll inflationary models, primordial fluctuations are nearly Gaussian and fully characterized by their power spectrum. However, many extensions of the standard inflationary scenario predict small but measurable departures from Gaussianity that contain valuable information about the particle content and interactions of the early Universe.

In this work we focus on the local type of primordial non-Gaussianity, in which the primordial gravitational potential is written as \citep{Gangui_1994}
\begin{align}
\Phi = \phi + f_\mathrm{NL}(\phi^2 - \langle \phi^2 \rangle),
\label{eq:png_pot}
\end{align}
where $\phi$ is a Gaussian random field and $f_\mathrm{NL}$ quantifies the amplitude of the non-Gaussian correction.

Local PNG produces a characteristic scale-dependent bias in the clustering of dark matter halos and galaxies on large scales \citep{Dalal_2008,Matarrese_2008,Slosar_2008} due to the coupling between long- and short-wavelength modes in the primordial potential \citep{gia-por-2010}. This effect enhances clustering at low wavenumbers in proportion to $f_\mathrm{NL}$, providing a powerful observational probe through large-scale structure surveys \citep{gia-por-car-2012}.

Current constraints from the cosmic microwave background are consistent with Gaussian initial conditions. The Planck satellite reports $f_\mathrm{NL} = -0.9 \pm 5.1$ \citep{PLANCK2018PNG}. LSS observations provide a complementary probe that is particularly sensitive on very large scales where the scale-dependent bias signal is strongest.

Constraining $f_\mathrm{NL}$ is therefore a key objective in modern cosmology. In this work we use PNG as a controlled test case to compare likelihood-based and simulation-based inference applied to summary statistics of the halo density field.

\subsection{Simulations}\label{sec:sims}
We use two sets of $N$-body simulations for different purposes. To train and validate the SBI pipeline, we use 1000 simulations from the publicly available \textsc{Quijote-PNG} suite \citep{Quijote_PNG}, in which $f_\mathrm{NL}$ is varied within the range $[-300,300]$ while all other cosmological parameters are kept fixed (the \texttt{latin\_hypercube\_LC} dataset). In addition, we use 2000 simulations from the \textsc{Quijote} suite \citep{Quijote} with $f_\mathrm{NL}=0$. These are used to estimate the covariance matrices for the LBI analysis and to perform frequentist validation of the inferred posteriors at fixed true value, as described in the remainder of the paper.

All simulations evolve $512^3$ dark matter particles in a periodic cubic box of volume $V = (1000\,h^{-1}\mathrm{Mpc})^3$. The cosmological parameters are fixed and consistent with \cite{PLANCK2018}: $\Omega_m=0.3175$, $\Omega_b=0.049$, $h=0.6711$, $n_s=0.9624$, $\sigma_8=0.834$, $w=-1.0$, and zero neutrino mass.

Halo catalogs (provided with the simulation suites) are identified using the Friends-of-Friends algorithm with standard linking length parameter $b=0.2$. In the following analysis we use halos with masses $M\geq3.2\times10^{13}\,h^{-1}\mathrm{M}_\odot$ at redshift $z=0$. This mass threshold corresponds to approximately 50 dark matter particles per halo.

To compute clustering statistics, we first construct the halo overdensity field
$\delta(\bm{x})$ for each halo catalog using the Piecewise Cubic
Spline (PCS) assignment scheme implemented in the \textsc{Pylians}
library\footnote{\url{https://pylians3.readthedocs.io/}} in combination
with grid interlacing. Grid-assignment effects are corrected by
deconvolving the PCS window function \citep{Sefusatti_2016}. We neglect redshift-space distortions, and the analysis is performed in real space.

\subsection{Power spectrum and bispectrum}

To extract statistical information from the halo density field, we consider the two- and three-point correlation functions in Fourier space, characterized by the power spectrum and the bispectrum. These statistics capture the Gaussian and leading non-Gaussian features of the field, respectively, and are standard probes of large-scale structure.

The power spectrum $P(k)$ and bispectrum $B(k_1, k_2, k_3 )$ are computed from the halo catalogs using the \textsc{PySpectrum} package.\footnote{\url{https://github.com/changhoonhahn/pySpectrum/}} The statistics are evaluated on a $360^3$ Fourier grid with bin width $\Delta k = 3k_f$, where $k_f = (2\pi/1000)\,h\,\mathrm{Mpc}^{-1}$ is the fundamental mode of the simulation box. For a bin of width $\Delta k$ centered on $k$, the power spectrum estimator is
\begin{align}
    \hat{P}(k)=\frac{1}{V\,N_k}
    \sum_{\bm{q}\in S_k}|\tilde\delta(\bm{q})|^2,
\end{align}
where $\tilde\delta(\bm{q})$ is the Fourier transform of the halo overdensity field $\delta(\bm{x})$, and $N_k$ is the number of discrete Fourier modes contained in the bin
 $S_k = \left\{\bm{q} \,|\, k-\Delta k/2 \leq |\bm{q}| \leq k + \Delta k/2\right\}$.
Similarly, the bispectrum estimator is
\begin{align}
\hat{B}(k_1,k_2,k_3)
=
\frac{1}{V\,N_{\rm tri}}
\sum_{\{\bm{q}_i\in S_{k_i}\}_{i=1}^{3}}
\tilde\delta(\bm{q}_1)\,
\tilde\delta(\bm{q}_2)\,
\tilde\delta(\bm{q}_3)\,
\delta^{\rm K}_{\bm{q}_1+\bm{q}_2+\bm{q}_3},
\end{align}
where $N_{\rm tri}$ is the number of distinct Fourier-space triangles satisfying the closure condition within the bins centered on $(k_1,k_2,k_3)$, and $\delta^{\rm K}$ denotes the Kronecker delta.

We restrict the analysis to the range $k \in [0.009, 0.141]\,h\,\mathrm{Mpc}^{-1}$, yielding 7 power spectrum bins. The bispectrum is measured for 62 triangle configurations defined by wave modes within the selected $k$-range and binning scheme, and subject to the conditions $k_1 \geq k_2 \geq k_3$ and $k_1 \leq k_2+k_3$.

\subsection{Wavelet scattering transform}

The WST \citep{wst_network} has recently emerged as a powerful summary statistic for LSS analyses. The transform itself maps the input density field  to a collection of scattering coefficients, which encode higher-order correlations beyond standard  $n$-point statistics and provide complementary information to the power spectrum \citep{Valogiannis_2022a,Valogiannis_2022b,Lemos_2024,Peron_2024}.

An $n$-th-order scattering coefficient is constructed by iterating $n$ convolutions with wavelet filters $\psi_{j\ell m}$, each followed by a pointwise nonlinear modulus, and finally raising the result to an integration power $q$ before averaging. In practice, one often works with rotationally invariant scattering coefficients, obtained by averaging over the angular index $m$, so that the resulting quantities do not depend on the orientation of the field. At zeroth order, this reduces to
\begin{align}
S^{(0)} = \left\langle \delta^q \right\rangle.
\end{align}
The first-order rotationally invariant coefficients involve a single convolution,
\begin{align}
    S^{(1)}_{j \ell} = \left\langle \left( \sum_{m=-\ell}^{\ell} \left| \delta \ast \psi_{j \ell m} \right|^2 \right)^{\frac{q}{2}} \right\rangle,
    \label{eq:s1}
\end{align}
where $j$ sets the smoothing scale and $\ell$ the angular resolution. Applying a second convolution yields the second-order coefficients. In full generality, these depend on two angular frequencies $\ell_1$ and $\ell_2$. In this work, we restrict to the case $\ell_1=\ell_2\equiv\ell$ and define
\begin{align}
    S^{(2)}_{j_1 j_2 \ell} = \left\langle \left( \sum_{m=-\ell}^{\ell} \left| \left( \sum_{m'=-\ell}^{\ell} \left| \delta \ast \psi_{j_1 \ell m'} \right|^2 \right)^{\frac{1}{2}} \ast \psi_{j_2 \ell m} \right|^2 \right)^{\frac{q}{2}} \right\rangle.
    \label{eq:S2def}
\end{align}
This restriction reduces the dimensionality of the coefficient space while retaining sensitivity to scale-dependent and angular features of the density field.

Although it is often stated in the cosmological literature that $n$-th-order WST coefficients depend on normalized moments of $\delta$ up to order $2^n$, this result strictly applies only to the original formulation \citep[see Section 4.1 of][]{mallat_2012}. This formulation does not coincide with the rotation-invariant definitions introduced above and commonly used in practice. We show in Appendix \ref{app:wst_ps} that the rotation-invariant WST considered here (at $q=2$) encodes information from the full hierarchy of correlation functions.

To compute the scattering coefficients, we first apply a sharp-$k$ band-pass filter to the halo density field, restricting the wavenumbers to the range $k \in [0.009,\,0.141]\,h\,\mathrm{Mpc}^{-1}$, consistent with the scales used for the power spectrum and bispectrum. This ensures that the WST does not access modes outside those contributing to $P$ and $B$, enabling a fair comparison of the information content of the different statistics while isolating large-scale contributions. The WST is then evaluated using the \textsc{Kymatio} package \citep{Andreux_20}. For three-dimensional fields, \textsc{Kymatio} employs solid harmonic wavelets 
\citep{scattering},
\begin{align}
    \psi_{j\ell m}(\bm{r}) = \frac{1}{(2 \pi)^{3/2} \sigma_j^3} e^{-r^2/2\sigma_j^2}\, \left(\frac{r}{\sigma_j}\right)^\ell\, Y_{\ell m}\left(\hat{\bm{r}}\right),
    \label{eq:solid_harmonic}
\end{align}
where $Y_{\ell m}$ denotes a spherical harmonic function,
$j \in [0, J]$, $\ell \in [0, L]$, $m \in [-\ell, \ell]$, and $\sigma_j = 2^j \sigma_0$. We adopt $J=6$, $L=4$, $\sigma_0=8.33 \, h^{-1}$ Mpc (corresponding to 3 grid cells), and integration power $q=2$.

For $q=2$, first-order WST coefficients depend only on the power spectrum and therefore contain no additional information beyond two-point statistics---see Eqs.~(\ref{eq:FTwavelet}) and (\ref{eq:s1power}). This equivalence explains the result of \cite{Peron_2024}, which finds that the power spectrum and first-order WST coefficients carry identical Fisher information on $f_\mathrm{NL}$, while second-order coefficients probe higher-order correlations.

The full set of WST coefficients is highly redundant. We therefore restrict the analysis to a subset of informative and weakly correlated coefficients, reducing the data vector to 41 elements without loss of constraining power. Details of the selection procedure are provided in Appendix~\ref{app:wst_ps}.

\subsection{Shot noise}\label{subsec:shot_noise}
We consider relatively rare and massive halos, with a mean number density $n \simeq 1.56 \times 10^{-4}\,h^3\,\mathrm{Mpc}^{-3}$ for $f_\mathrm{NL}=0$, which varies mildly (less than $1\%$) with $f_\mathrm{NL}$. Because our analysis is restricted to large scales ($k\le0.141\,h\,\mathrm{Mpc}^{-1}$), the corresponding shot noise is expected to be subdominant for the power spectrum but non-negligible for some bispectrum configurations.

Since we do not have an explicit model for the systematic shot-noise contribution to the WST coefficients, we do not apply shot-noise corrections to the power spectrum or bispectrum. This ensures a fair comparison between summary statistics, as the PNG signal correlates nontrivially with shot noise. While an analytic expression for the shot-noise contribution to the first-order WST coefficients can be derived (see Appendix~\ref{sec:WST_shot_noise}), no such model is currently available for the second-order coefficients.

The mild dependence of the halo number density on $f_\mathrm{NL}$ implies that shot noise is, in principle, parameter-dependent and could introduce additional sensitivity to PNG in the summary statistics. However, we have explicitly verified that this
effect is negligible by performing inference using $P$ and $B$ both with and without shot-noise corrections, finding nearly identical constraints on $f_\mathrm{NL}$. This indicates that shot noise does not significantly affect the extracted PNG information in the regime considered here.

\subsection{Simulation-based inference}\label{subsec:sbi_implement}

We perform SBI using the \textsc{sbi} package \citep{BoeltsDeistler_sbi_2025}. In particular, we make use of their NRE\_C module; a CNRE implementation of the method introduced in \citep{cnre_paperb}. The model is trained on 800 simulations from the Latin hypercube dataset, with 200 simulations used for validation. For testing, we use the same 1000 fiducial simulations employed in the likelihood-based analysis, enabling direct realization-by-realization comparisons of posterior estimates.

The number of simulations required for SBI depends on the dimensionality of the parameter space and data vector, as well as the complexity of the posterior. Previous work \citep{Bairagi_2025,Homer_2025} suggests that many SBI applications are likely under-trained and that increasing the number of simulations is generally more beneficial than increasing model complexity. In our setup (one parameter and a data vector of size $\mathcal{O}(10$--$100)$), we find that posterior estimates are stable for training set sizes between 600 and 1000 simulations, motivating our choice of 800 training simulations.

We calibrate the training hyperparameters using two approaches: (i) a diagnostic-based procedure relying on the SBC rank test and TARP coverage test (hereafter SBI$_\mathrm{sbc}$), and (ii) direct optimization of the cross-entropy loss (hereafter SBI$_\mathrm{loss}$). Details are given in Appendix~\ref{appndx:diagnostics}, where we also show the SBC and TARP test results for our SBI$_\mathrm{sbc}$ models.

Since $P$ and $B$ differ significantly in scale and are correlated, we tested three data-combination strategies: using the raw data, standardizing to zero mean and unit variance, and pre-whitening. While the qualitative behavior is unchanged across all preprocessing strategies, we observe a modest change for the $\mathrm{SBI}_\mathrm{loss}$ model trained on the rescaled data vector. This is the version of the $\mathrm{SBI}_\mathrm{loss}$ model considered in the following.

\subsection{Likelihood-based inference}\label{subsec:lbi_implement}

\subsubsection{Surrogate models}
For likelihood-based inference, we constructed a parametric model for each summary statistic as a function of $f_\mathrm{NL}$. Since no accurate analytical model is available for the statistics considered here, we instead adopted simple surrogate models motivated by perturbation theory.

Perturbative calculations for the power spectrum of biased tracers indicate that, at next-to-leading order, the deterministic contribution can be expressed as a combination of terms proportional to a constant, $f_\mathrm{NL}$, and $(f_\mathrm{NL})^2$. Stochastic contributions (shot noise) introduce corrections scaling as $\bar{n}^{-1}$, where $\bar{n}$ is the tracer number density. From the simulations, we find that, over the range explored here, the mean halo density is well described by $\bar{n} \simeq \bar{n}_0\, (1 + \alpha f_\mathrm{NL})$, with a maximum deviation of $\sim 1\%$ at $|f_\mathrm{NL}| \simeq 300$. It is therefore natural to expand
\begin{align}
\frac{1}{\bar{n}}(f_\mathrm{NL})
\simeq
\frac{1}{\bar{n}_0}
\left[
1 - \alpha f_\mathrm{NL}
+ \alpha^2 (f_\mathrm{NL})^2
+ \cdots
\right].
\end{align}
In addition, the coefficients multiplying $\bar{n}^{-1}$ in the shot-noise terms can themselves depend linearly on
$f_\mathrm{NL}$ \citep{Hamaus+2011}. Together, these considerations suggest that the large-scale dependence of
the halo power spectrum on $f_\mathrm{NL}$ can be accurately approximated by a low-order polynomial at fixed wavenumber.

A similar argument applies to the bispectrum. At leading order, the deterministic contribution contains terms proportional to a constant, $f_\mathrm{NL}$, $(f_\mathrm{NL})^2$, and $(f_\mathrm{NL})^3$ \citep[e.g.,][]{Baldauf_2011}. Additional contributions arise from shot noise, including terms scaling as $\bar{n}^{-1}$ and $\bar{n}^{-2}$, whose coefficients may also depend on $f_\mathrm{NL}$. This again motivates modeling the dependence of the halo bispectrum on $f_\mathrm{NL}$ at fixed triangle configuration with a low-order polynomial.

We assume that the same reasoning applies to the WST coefficients, whose leading contributions are dominated by the power spectrum and trispectrum (see Appendix~\ref{appndx:wst}). This motivates the expectation that their dependence on $f_\mathrm{NL}$ is likewise smooth over the parameter range considered here.

In practice, we fit an $n^\mathrm{th}$-order polynomial with $n = 2, 3, 4$ independently to each component of the data vector using least-squares regression on the 1000 \texttt{Latin\_hypercube\_LC} simulations. We find that all choices provide very similar fits, with differences at the level of at most $\sim 1\%$ of the rms scatter across simulations. The largest deviation is observed for the WST, reaching $1.4\%$ for a single coefficient. Thus, our main conclusions are insensitive to the choice of polynomial order. In the remainder of this work, we present results obtained with $n=2$, unless stated otherwise.

\subsubsection{Likelihood function}

A common practice in cosmology is to model the likelihood as Gaussian. In this case, the information content of the data is fully characterized by its mean and covariance matrix $\Sigma$. The log-likelihood then takes the form
\begin{align}
\log p(x|\theta) = -\frac{1}{2} D^\top \Sigma^{-1} D + \mathrm{const.},
\end{align}
where $D = x_{\mathrm{obs}} - x_{\mathrm{model}}$.

In principle, the covariance matrix entering the Gaussian likelihood depends on the model parameters, but in cosmological applications it is typically evaluated at a fiducial model and held fixed in order to
simplify the inference. In our case, we estimate the sample covariance $S$ from $n_\mathrm{s}=800$ realizations of the \textsc{Quijote} suite and use a separate set of 1000 simulations for testing. The covariance is thus estimated from the same number of simulations used to train the SBI, ensuring that finite-sample effects are unlikely to drive the observed discrepancies, provided that the covariance matrix depends only weakly on $f_\mathrm{NL}$. We have verified that increasing the number of simulations used for covariance estimation to 1000 leaves our results unchanged.

The statistical properties of the sample covariance have long been studied in multivariate statistics. For Gaussian-distributed data, $S$ follows a Wishart distribution \citep{Wishart1928}, and its inverse provides a biased estimator of the true precision matrix. In particular, the frequentist expectation value satisfies $\overline{S^{-1}} \neq \Sigma^{-1}$, where the overbar denotes
an average over repeated realizations of $n_\mathrm{s}$ samples \citep[e.g.,][]{Muirhead1982, Anderson_statistics_2003}. This bias is commonly corrected by rescaling the inverse sample covariance using the Anderson--Hartlap factor \citep{Hartlap_2006},
\begin{align}
\Sigma^{-1} \;\rightarrow\; f_\mathrm{AH}\, S^{-1}, \qquad
f_\mathrm{AH} = \frac{n_\mathrm{s} - n_\mathrm{D} - 2}{n_\mathrm{s} - 1},
\end{align}
where $n_\mathrm{D}$ is the dimension of the data vector.

In addition, uncertainty in the estimated covariance matrix propagates into parameter constraints, leading to an underestimation of posterior variances if unaccounted for. The Dodelson--Schneider correction accounts for the increased scatter of the maximum a posteriori (MAP) estimator induced by noise in the estimated covariance matrix, and amounts to a rescaling of the parameter covariance by a factor
\begin{align}
f_\mathrm{DS} =
1 + \frac{(n_\mathrm{D} - n_\theta)(n_\mathrm{s} - n_\mathrm{D} - 2)}
{(n_\mathrm{s} - n_\mathrm{D} - 1)(n_\mathrm{s} - n_\mathrm{D} - 4)},
\end{align}
where $n_\theta$ is the number of inferred parameters \citep{Dodelson_2013}. This correction has a frequentist interpretation, as it is derived from the variance of parameter estimates across repeated realizations.

A fully Bayesian treatment would instead account for uncertainty in the covariance matrix at the likelihood level by marginalizing over it. This leads to a multivariate $t$-distribution with heavier tails \citep{Sellentin-Heavens_2016}. In general, Bayesian credible intervals obtained in this way are not guaranteed to have exact frequentist coverage, i.e., the probability that they contain the true parameter value across realizations (at fixed truth) does not necessarily match their nominal credibility.

\citet{Percival_2021} propose an alternative approach in which the prior on the covariance matrix is chosen as a "probability-matching prior", ensuring that Bayesian credible intervals recover the correct frequentist coverage. In practice, a commonly used approximation to this full treatment is obtained by effectively rescaling the covariance matrix in the Gaussian likelihood \citep[see also][]{Percival_2014, Friedrich-Eiffler_2018}
\begin{align}
\Sigma \;\rightarrow\; f_\mathrm{P}\, S, \qquad
f_\mathrm{P} =
\frac{(n_\mathrm{s}-1)\,f_\mathrm{DS}}{n_\mathrm{s}-n_\mathrm{D}+n_\theta-1}
\;\simeq\; \frac{f_\mathrm{DS}}{f_\mathrm{AH}}.
\end{align}
We adopt this simplified approximation in our analysis.

These corrections are now standard in LSS analyses. We note, however, that they rely on the assumption of Gaussian-distributed
summary statistics and Wishart-distributed sample covariances. In cases where the data vector exhibits significant non-Gaussianity these treatments should be regarded as approximate, and their validity may need to be reassessed.

To obtain posterior distributions in the LBI approach, we first evaluate the likelihood on a dense grid of 1000 uniformly spaced values of $f_\mathrm{NL}$
spanning the prior range, and then normalize it by the marginal likelihood (model evidence) computed numerically.

\section{Results}\label{sec:results}

Standard calibration diagnostics suggest that the SBI posteriors are well behaved. However, a realization-level comparison with likelihood-based inference reveals systematic differences that are not captured by those diagnostics.

We obtain posterior estimates for 1000 reserved test simulations drawn from the 2000 fiducial \textsc{Quijote} realizations with $f_\mathrm{NL}=0$, using both the posterior distributions obtained with our LBI approach and our trained SBI posterior estimator based on CNRE. In what follows, we first assess the Gaussian-likelihood assumption underlying the LBI analysis, and then compare the posterior distributions obtained with the two methods in a frequentist fashion across the test ensemble.

\subsection{Testing the Gaussian likelihood approximation}\label{subsec:testing_likelihood_main}

Whether the LBI results can be used as a meaningful reference for the SBI posteriors depends on the validity of the assumptions entering the LBI pipeline. In particular, we assume that on the large scales considered here, $k\in[0.009,0.141]\,h\,\mathrm{Mpc}^{-1}$, the summary statistics retain an approximately Gaussian likelihood.

Standard visual diagnostics such as Q--Q plots of Mahalanobis distances provide only limited information and are not sufficiently sensitive. We therefore rely on a suite of multivariate normality tests to assess this assumption quantitatively. Specifically, we apply six tests probing complementary aspects of multivariate normality to each summary statistic individually, as well as to their combinations, using the R \textsc{mnt} library \citep{EbnerHenze2020}.

First, we use Mardia's skewness and kurtosis tests
\citep{mardia1970}, which explicitly probe the third- and fourth-order
moments of the distribution. Mardia skewness is primarily sensitive to
asymmetries and nonlinear correlations, while Mardia kurtosis tests the
radial structure of the data, i.e., the distribution of Mahalanobis distances from the mean,  providing a direct diagnostic of
deviations in the tails relative to a Gaussian expectation.

Second, we use the distance-based Henze--Zirkler (HZ) test \citep{henze_zirkler} and the Székely--Rizzo (SR) energy test \citep{szekely2005}, which quantify the deviation of the empirical distribution from a multivariate Gaussian in a global sense. The HZ test is based on a comparison between the empirical characteristic function of the data and that of a multivariate Gaussian
distribution with matching mean and covariance. It effectively measures a weighted $L^2$ distance between these two functions, making it sensitive to a broad class of deviations from Gaussianity across all scales. The SR energy test, on the other hand, is formulated in terms of pairwise Euclidean distances between data points. It compares the average distances within the sample to those expected under a Gaussian distribution, thereby providing a non-parametric and rotation-invariant measure of discrepancy. This test is particularly sensitive to global features such as clustering, multimodality, or deviations in dispersion. Both tests are therefore sensitive to a wide range of departures from Gaussianity, including skewness, kurtosis, and more general structural differences, although they do not directly identify which specific feature is responsible for a detected deviation.

Finally, we apply the two Dörr--Ebner--Henze tests, denoted DEHU and DEHT in the \textsc{mnt} library, which are among the most powerful omnibus tests of multivariate normality. These tests are based on distinct characterizations of the Gaussian distribution: DEHU relies on a partial differential equation uniquely satisfied by the Gaussian \citep{DEHU_test}, while DEHT is constructed from an expansion in Hermite functions \citep{DEHT_test}. Because they are grounded in fundamentally different properties of the Gaussian distribution, they are sensitive to subtle and structured deviations that may not be captured by moment-based or distance-based tests alone. Taken together, these tests allow us to disentangle whether deviations
from Gaussianity arise from asymmetry, tail behavior, or more general
structural effects.

\begin{table}
\caption{Summary of multivariate normality test results for the individual and combined summary statistics}.
\label{tab:mnt_mvn_results}
\centering

\begin{tabular*}{\columnwidth}{@{\extracolsep{\fill}}lccccc@{}}
\hline\hline
Test & $P$ & $B$ & $P+B$ & WST & $P+B+$WST \\
\hline
Mardia $\gamma$ & \cmark & \xmark & \xmark & \xmark & \xmark \\
Mardia $\kappa$ & \cmark & \cmark & \cmark & \xmark & \cmark \\
HZ              & \cmark & \xmark & \xmark & \cmark & \cmark \\
SR Energy       & \cmark & \xmark & \xmark & \cmark & \cmark \\
DEHT            & \cmark & \xmark & \xmark & \xmark & \xmark \\
DEHU            & \cmark & \xmark & \xmark & \xmark & \xmark \\
\hline
\end{tabular*}
\tablefoot{The checkmark indicates that a certain test is passed at $95\%$ confidence level, and a cross indicates otherwise.}
\end{table}

The results are summarized in Table~\ref{tab:mnt_mvn_results}. Among the summary statistics considered here, $P$ is the only one that consistently passes all tests. In contrast, $B$ and $P+B$ pass only Mardia kurtosis while failing all other diagnostics, indicating that their departure from Gaussianity is not primarily driven by anomalous kurtosis, but instead reflects more general structural features such as asymmetry or nonlinear dependence among components.

The WST coefficients pass the distance-based HZ and SR tests, suggesting no strong global deviation from Gaussianity. However, they fail Mardia and DEH tests, revealing non-Gaussian features in higher-order moments, including asymmetry and deviations in the tails. This implies that a Gaussian likelihood may provide a reasonable approximation for the bulk of the distribution, but may not accurately capture its higher-order structure.

The combined data vector $P+B+\mathrm{WST}$ passes the HZ and SR tests, as well as Mardia kurtosis, indicating that its distribution is globally close to Gaussian and exhibits approximately Gaussian tail behavior. However, the failure of Mardia skewness and both DEH tests reveals the presence of asymmetry and subtler structured deviations
from Gaussianity. This suggests that, while the Gaussian approximation captures the overall dispersion and radial structure of the data, it does not fully describe its multivariate dependence structure.

Overall, these results indicate that all data vectors except $P$ deviate from multivariate normality, with differences primarily arising from higher-order structure rather than from gross departures in the bulk distribution.

This implies that the Gaussian-likelihood assumption underlying LBI is most reliable for $P$, making the corresponding posterior estimates the most defensible reference point for comparison with SBI. For the remaining statistics, however, non-Gaussianity in the data vector may itself induce differences between the LBI and SBI posteriors. This caveat should be kept in mind in the comparisons below.

\subsection{Comparing LBI and SBI posteriors}\label{subsec:comparison}

\begin{figure*}
    \centering
    \includegraphics[width=0.94\linewidth]{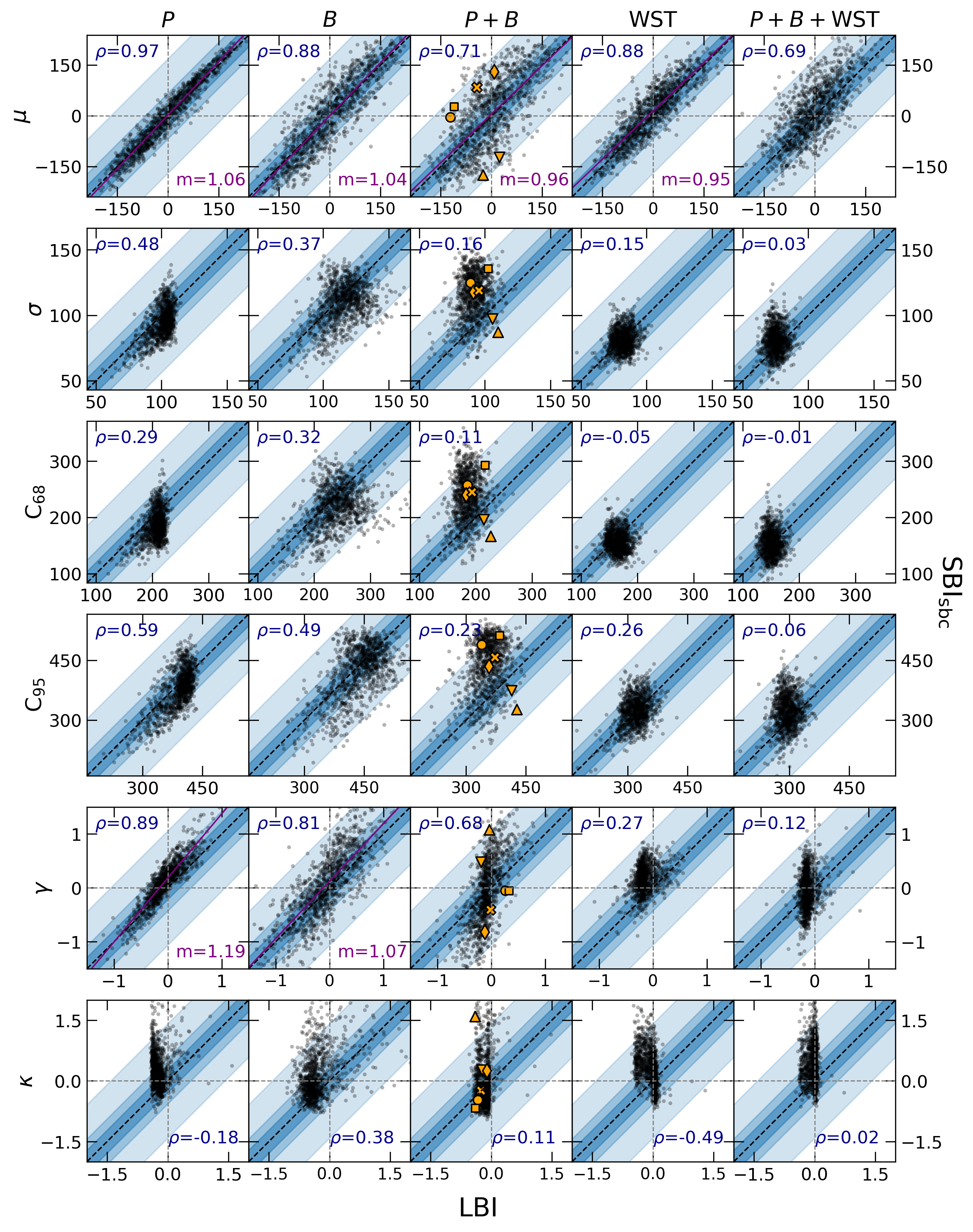}
    \caption{
Comparison of posterior properties obtained with LBI ($x$-axis) and SBI$_\mathrm{sbc}$ ($y$-axis) across 1000 test realizations with $f_\mathrm{NL}=0$. Each black point corresponds to one realization. The six realizations shown in Fig.~\ref{fig:p_posts_comp} are highlighted with orange symbols. From top to bottom, the panels show the posterior mean $\mu$, standard deviation $\sigma$, 68\% and 95\% credible interval widths, skewness $\gamma$, and excess kurtosis $\kappa$. The black dashed line indicates equality ($y=x$), while the shaded regions denote relative deviations of 5\%, 10\%, and 25\%. Horizontal and vertical gray dashed lines indicate zero values. The Spearman rank correlation coefficient $\rho$ is reported in the top-left corner. If $\rho > 0.7$, a linear regression is also shown (purple line), with its slope $m$ indicated in the bottom-right corner.
}
  
    \label{fig:big_plot_main}
\end{figure*}

\begin{figure}
    \centering
    \resizebox{\hsize}{!}{\includegraphics{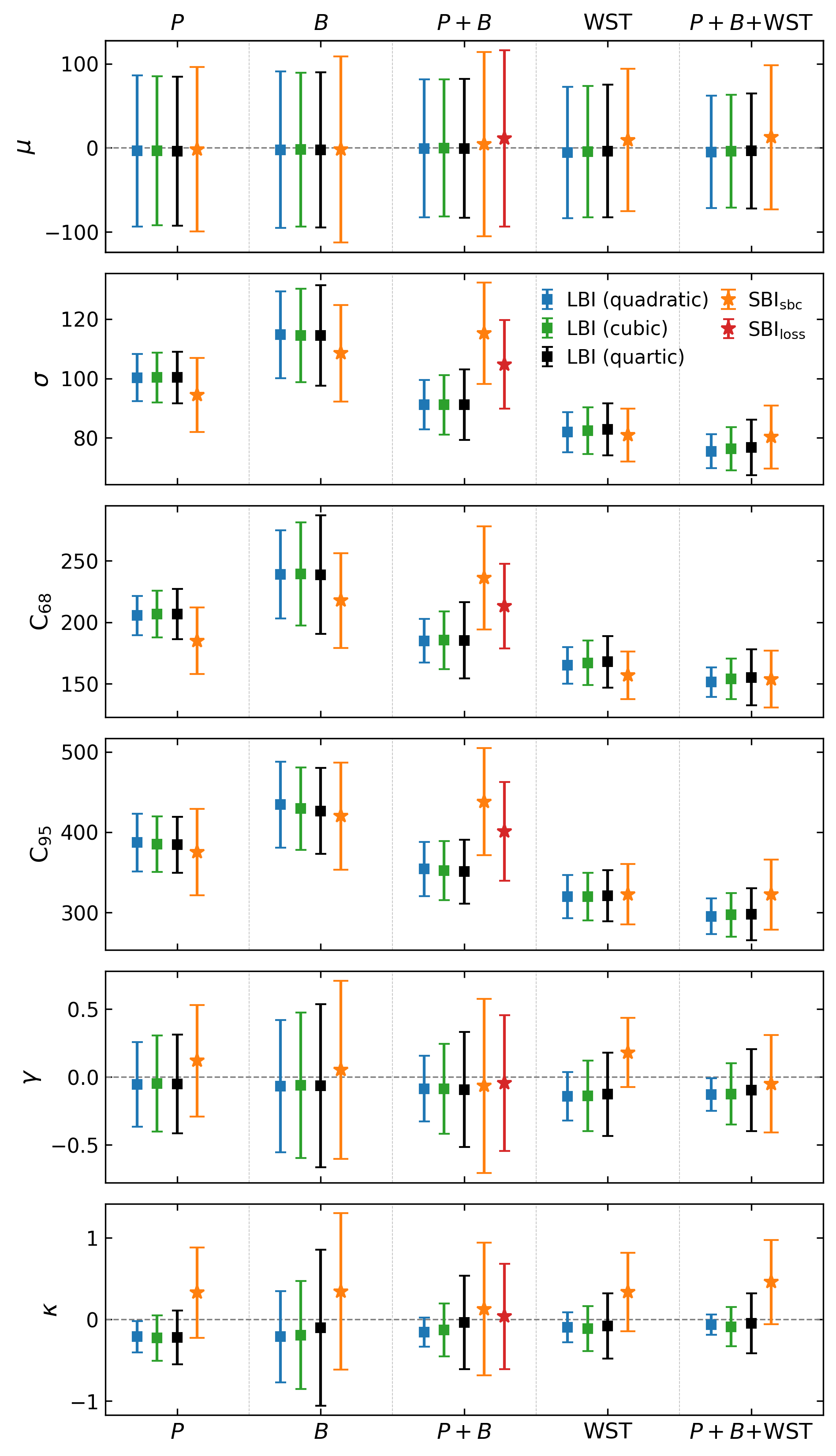}}
    \caption{Summary statistics (mean, standard deviation, 68\% and 95\% credible interval widths, skewness, and excess kurtosis) of 1000 posterior estimates from SBI (CNRE) and LBI, evaluated on realizations with $f_\mathrm{NL}=0$. LBI results are shown for all three polynomial surrogate models. Error bars indicate the rms scatter across realizations.}
    \label{fig:posts_stats}
\end{figure}

We compare posterior estimates obtained from LBI and SBI\footnote{We find that the two SBI implementations (SBI$_\mathrm{sbc}$ and SBI$_\mathrm{loss}$) yield nearly identical results in all cases except for the $P+B$ data vector. For clarity, we therefore present results obtained with SBI$_\mathrm{sbc}$ throughout, and discuss SBI$_\mathrm{loss}$ only for the $P+B$ case where differences arise.} for five data vectors: $P$, $B$, $P+B$, WST, and $P+B+$WST. For each of the 1000 test realizations, we examine several properties of the posterior distribution: the mean $\mu$, standard deviation $\sigma$, widths of the 68\% and 95\% credibility intervals, skewness $\gamma$, and excess kurtosis $\kappa$. The LBI posterior is evaluated on a grid of 1000 $f_{\rm NL}$
values, while the SBI posterior is represented by 5000 samples generated via rejection sampling.

\subsubsection{Comparison of posterior moments across realizations}

Figure~\ref{fig:big_plot_main} shows the realization-by-realization comparison of these quantities. For the power spectrum, the comparison between SBI and LBI posterior estimates reveals a clear hierarchy in the agreement of statistical moments. The rank correlation coefficient between the two methods is 0.97 for the posterior mean, 0.48 for the standard deviation, 0.89 for the skewness, and $-0.18$ for the kurtosis. In particular, $\mu$ and $\gamma$ agree at the
$\lesssim 10\%$ level for most realizations, indicating that both methods recover the location and asymmetry of the posterior with high consistency. In contrast, $\sigma$, credible interval widths, and $\kappa$ show significantly larger realization-to-realization scatter for SBI, while LBI estimates cluster around typical values. Although deviations are generally below $\sim 25\%$, qualitative differences are evident: LBI posteriors tend to be slightly platykurtic, whereas SBI posteriors are predominantly leptokurtic. This highlights a systematic difference in how the two methods describe the tails of the posterior distribution, with SBI posteriors tending to exhibit broader tails and more pronounced central peaks.

These trends persist for all other summary statistics and their combinations, although with increased scatter. The similarity of this behavior across different statistics---including the power spectrum, for which the Gaussian-likelihood approximation is expected to be accurate---suggests that the observed features are not primarily driven by the non-Gaussianity of the data-generating process. Instead, they point to systematic differences in how LBI and SBI respond to sampling noise and encode posterior structure.

Figure~\ref{fig:posts_stats} shows the corresponding ensemble averages and scatters of the posterior means, standard deviations, and credibility-interval widths. The posterior means from the two methods are nearly identical on average for all data vectors and are consistently very close to the true parameter value. The standard deviations and interval widths also agree reasonably well on average in most cases, although SBI posteriors exhibit a larger realization-to-realization scatter.\footnote{Since $P$ and $B$ differ significantly in scale and are correlated, we tested three data-combination strategies: using the raw data, standardizing to zero mean and unit variance, and pre-whitening. While the qualitative behavior is unchanged across all preprocessing
strategies, we observe a modest change for the $\mathrm{SBI}_\mathrm{loss}$ model trained on the rescaled data vector. This is the version of the $\mathrm{SBI}_\mathrm{loss}$ model considered in the following.} This indicates that SBI tends to produce more variable uncertainty estimates, especially for weakly constraining realizations.

Results for the cubic and quartic LBI models are nearly identical to the quadratic case, with only a modest increase in the scatter of higher-order posterior moments. In the idealized case of a linear model with a Gaussian likelihood, the posterior is Gaussian (up to truncation by the prior bounds). Although our forward model is formally nonlinear due to the polynomial parametrization, the small differences between polynomial orders indicate that nonlinear contributions are subdominant over the parameter range explored. As a result, the effective dependence on the parameters is close to linear, the likelihood remains approximately quadratic, and the inferred parameters are largely insensitive to small modeling variations.

For $P$, $B$, and WST, the SBI posteriors are on average slightly more peaked than the LBI posteriors, with correspondingly smaller values of $\sigma$. The main exception is the combined $P+B$ data vector, for which the SBI posteriors are systematically broader. This trend persists, though more weakly, when the WST coefficients are added to the combination.

\subsubsection{Conditional coverage}

In contrast to the (marginal) coverage under the prior predictive distribution, one can define the (conditional) coverage at fixed parameter value, corresponding to the standard frequentist notion of coverage under repeated sampling at fixed truth. It is important to note that coverage evaluated at fixed parameter value does not, in general, coincide with the nominal credibility level, even for an exact posterior. A match is obtained only under special conditions. Each of the following is sufficient to ensure agreement: (i) adopting probability-matching priors, (ii) the asymptotic regime regulated by the Bernstein--von Mises theorem (large data, regular model), and (iii) for approximately Gaussian posteriors. Deviations from the nominal level can thus arise from finite-sample effects or non-Gaussianity. Therefore, in this context, coverage should not be interpreted as an absolute measure of calibration but as a metric to measure relative differences between inference methods. Table~\ref{tab:coverage} reports the fractional coverage of the true value, $f_\mathrm{NL}=0$, within the 68\% and 95\% credible intervals. For LBI, the credible intervals indicate mild over-coverage relative to the nominal levels but are still broadly consistent with approximately Gaussian behavior in the bulk of the posterior. The combination of under-coverage at the 68\% level and nearly correct coverage at the 95\% level indicates that the SBI posteriors for $P$, $B$, and $P+B$ are more concentrated around their peak while compensating with broader tails. This behavior is consistent with a mis-specification of the posterior shape, in which the central region is overconfident while the tails are over-dispersed. Finally, when the WST measurements are considered, the coverage indicates that the posterior is systematically under-dispersed, with both the core and the tails underestimated, leading to uniformly over-confident constraints with respect to LBI.

\begin{table}
\caption{Fractional coverage of posterior estimates (LBI and SBI$_\mathrm{sbc}$) for $f_\mathrm{NL}^\mathrm{true}=0$ within 68\% and 95\% credible intervals across 1000 test samples.}
\label{tab:coverage}
\centering
\begin{tabular*}{\columnwidth}{@{\extracolsep{\fill}} l cccc}
\hline\hline
\noalign{\smallskip}
\multirow{2}{*}{Statistic} & \multicolumn{2}{c}{LBI} & \multicolumn{2}{c}{SBI} \\
\noalign{\smallskip}
 & $f_{68}$ & $f_{95}$ & $f_{68}$ & $f_{95}$ \\
\noalign{\smallskip}
\hline
\noalign{\smallskip}
$P$       & 0.71 & 0.97 & 0.58 & 0.93 \\
$B$       & 0.72 & 0.99 & 0.60 & 0.95 \\
$P+B$     & 0.71 & 0.97 & 0.62 & 0.96 \\
WST     & 0.70 & 0.97 & 0.59 & 0.91 \\
$P+B$+WST & 0.73 & 0.98 & 0.60 & 0.91 \\
\noalign{\smallskip}
\hline
\end{tabular*}
\end{table}

\subsubsection{Posterior discrepancy metrics}
To quantify differences between posterior distributions at the realization level, we employ both the Kullback--Leibler divergence (KLD) and the Wasserstein distance. These two measures capture complementary aspects of the discrepancy between distributions. The KLD is defined as
\begin{align}
D_{\mathrm{KL}}(P \,\|\, Q)
= \int P(\theta)\,
\log \frac{P(\theta)}{Q(\theta)} \,\mathrm{d}\theta,
\end{align}
and quantifies the difference in probability mass assignment. It is particularly sensitive to mismatches in the tails and to deviations in the overall posterior shape, providing a stringent diagnostic of differences that are not reflected in low-order summary statistics.

In contrast, the Wasserstein distance measures the minimal cost of transporting one distribution into the other. In one dimension, the $p$-Wasserstein distance can be written as
\begin{align}
W_p(P, Q) =
\left(
\int_0^1 \left| F_P^{-1}(u) - F_Q^{-1}(u) \right|^p \,\mathrm{d}u
\right)^{1/p},
\end{align}
where $F^{-1}$ denotes the inverse cumulative distribution function and $p$ denotes the order of the Wasserstein distance, which controls how differences between the two distributions are weighted, with larger values of $p$ placing greater emphasis on larger discrepancies. This expression shows that the Wasserstein distance corresponds to an average difference between the quantiles of the two distributions, and is therefore directly sensitive to differences in location and spread, with a clear interpretation in parameter units.

For Gaussian distributions, the $W_2$ distance admits a simple closed-form expression in terms of the mean and covariance. In particular, for two one-dimensional Gaussians $P = \mathcal{N}(\mu_1, \sigma_1^2)$ and $Q = \mathcal{N}(\mu_2, \sigma_2^2)$, the $W_2$ distance reduces to
\begin{align}
W_2^2(P, Q) = (\mu_1 - \mu_2)^2 + (\sigma_1 - \sigma_2)^2,
\end{align}
while other choices of $p$ generally do not yield comparably compact formulas. This provides useful intuition for its behavior when comparing posterior distributions.

By combining these two metrics, we are able to distinguish between discrepancies arising from shifts in the bulk of the posterior and those originating from differences in its tails or overall probability structure.

\begin{figure}
    \centering
    \resizebox{\hsize}{!}{\includegraphics{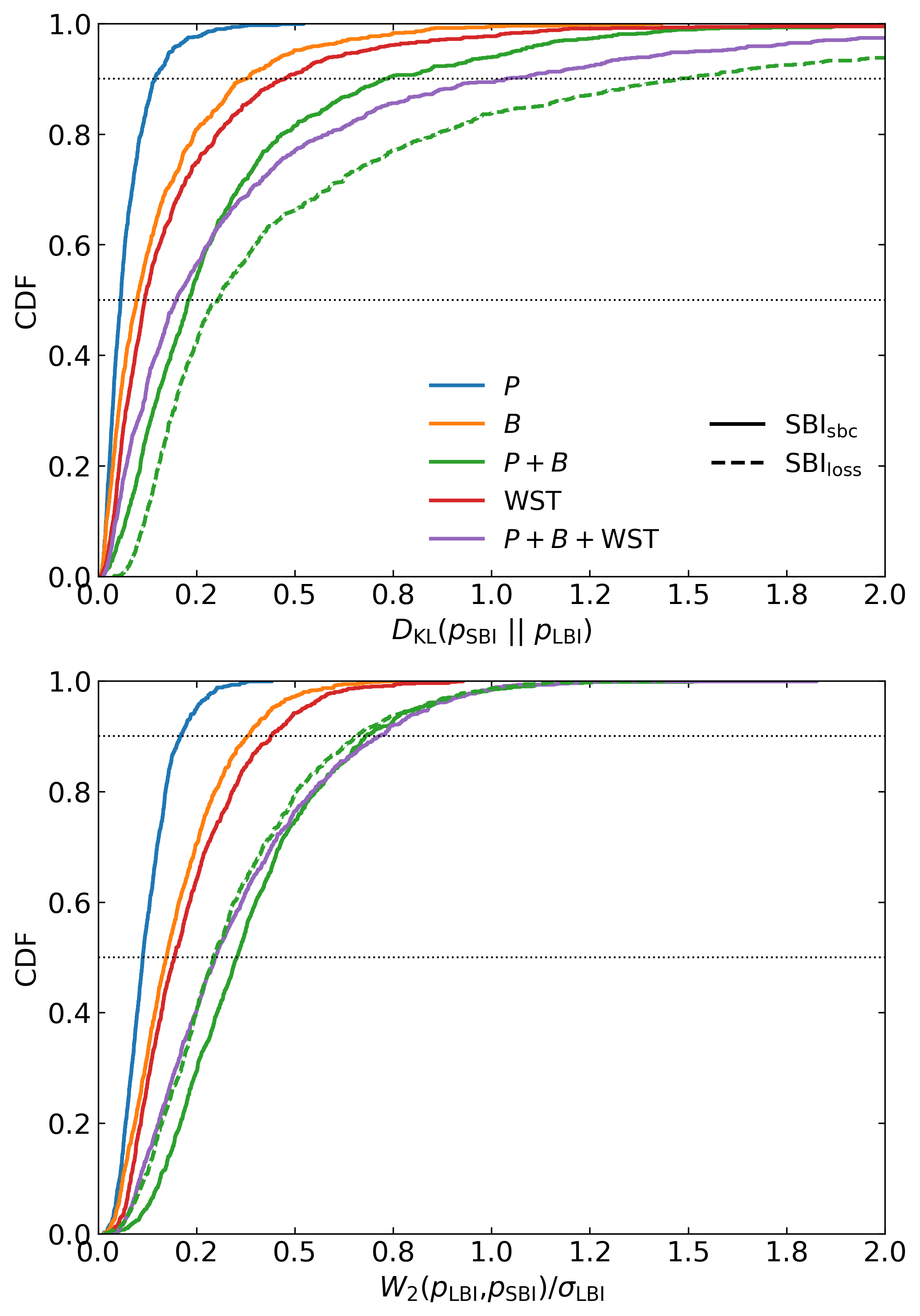}}
   \caption{CDF of the KLD and 2-Wasserstein distance between SBI and LBI posterior estimates for 1000 test realizations with $f_\mathrm{NL}=0$. The $W_2$ distance is normalized by $\sigma_\mathrm{LBI}$, the rms width of the LBI posterior, to provide a dimensionless measure of the discrepancy relative to the posterior width.  For comparison, the dashed line shows results for $\mathrm{SBI}_\mathrm{loss}$ in the $P+B$ case.}
   \label{fig:kld_w2}
\end{figure}
Figure~\ref{fig:kld_w2} shows the cumulative distribution functions (CDFs) of the KLD and the 2-Wasserstein distance between SBI and LBI posteriors across the test set. The CDF representation highlights both the typical level of agreement and the extent of realization-to-realization variability.

For the power spectrum, the distributions are sharply peaked at small distances, indicating very good agreement between the two inference methods. The bispectrum and WST coefficients exhibit broader distributions, with a non-negligible fraction of realizations showing moderate discrepancies. The combined data vectors, particularly $P+B$ and $P+B+\mathrm{WST}$, display the largest spread, reflecting increased sensitivity to differences in posterior shape as additional summary statistics are included, leading to higher dimensional and correlated data vectors. Notably, for the $P+B$ combination, $\mathrm{SBI}_\mathrm{loss}$ yields smaller values in the $W_2$ CDF than $\mathrm{SBI}_\mathrm{sbc}$, while exhibiting larger KLD values. This highlights that the two metrics probe different aspects of the discrepancy between posterior distributions.

\begin{figure}
    \centering
    \resizebox{\hsize}{!}{\includegraphics{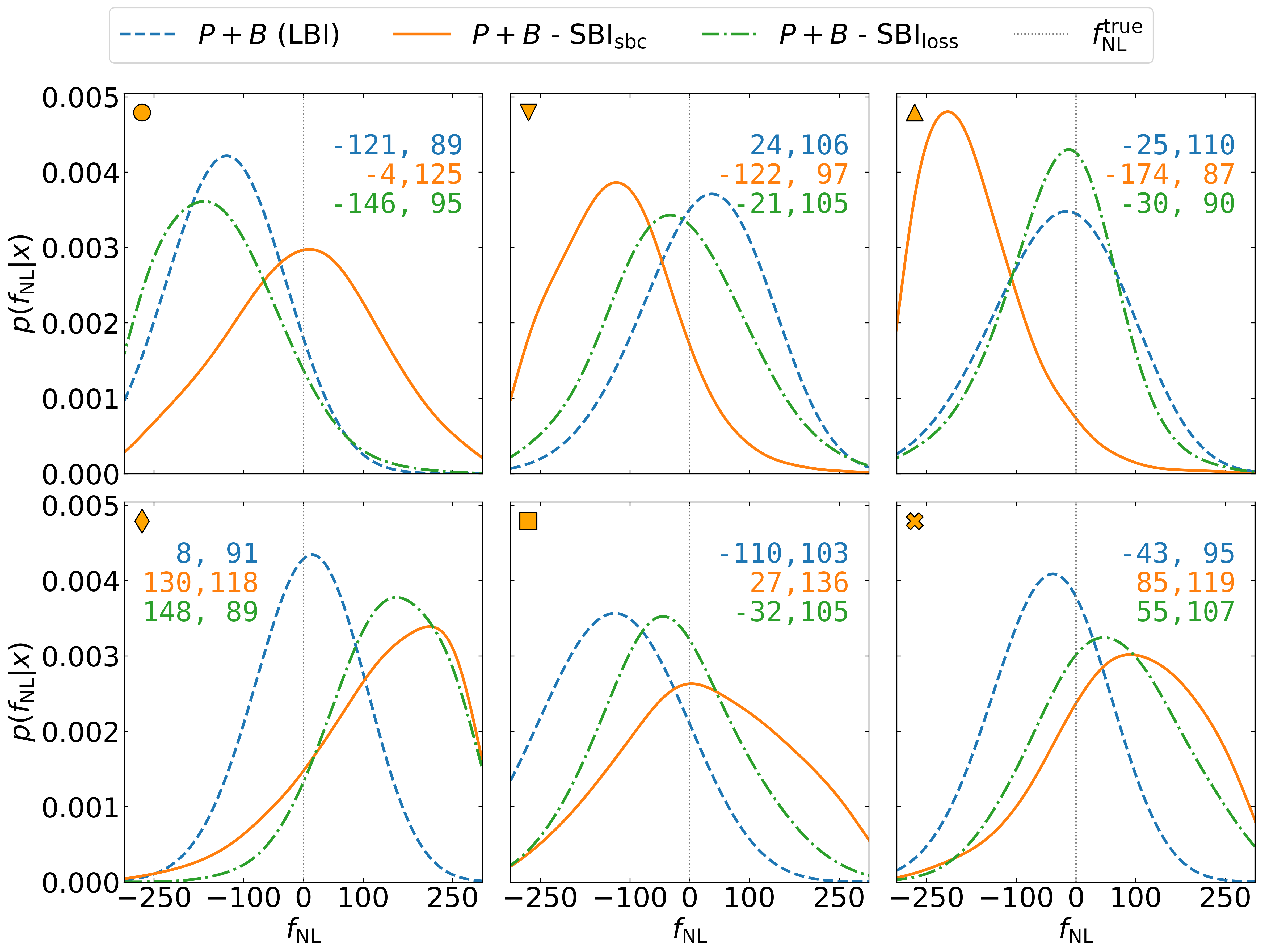}}
    \caption{Comparison of LBI (blue), SBI$_\mathrm{sbc}$ (orange), and SBI$_\mathrm{loss}$ (green) posteriors for six  test realizations with $f_\mathrm{NL}=0$ using $P+B$.
    These cases correspond to normalized 2-Wasserstein distances between the LBI and SBI$_\mathrm{sbc}$ posteriors close to the 90th percentile of the distribution in Fig.~\ref{fig:kld_w2}, and exhibit noticeable shifts and shape differences between the three
    methods. Markers match the corresponding realizations in Fig.~\ref{fig:big_plot_main}. Each panel reports the mean and standard deviation of the posterior as an ordered pair ($\mu$,$\sigma$) in the color corresponding to each model.}

    \label{fig:p_posts_comp}
\end{figure}

For these combined statistics, more than 50\% of the realizations have a 2-Wasserstein distance larger than the width of the corresponding LBI posterior, indicating substantial differences that can impact the inferred constraints on a per-realization basis. This is illustrated in Fig.~\ref{fig:p_posts_comp}, which shows six representative examples with 2-Wasserstein distances close to the 90th percentile of the distribution. In these cases, the SBI$_\mathrm{sbc}$, SBI$_\mathrm{loss}$, and LBI posteriors are visibly displaced relative to each other.

\subsection{Underconfident SBI posteriors for the combined P+B data vector}

Building on the discrepancies identified above, the most striking effect appears for the combined $P+B$ data vector, as shown in Figs.~\ref{fig:big_plot_main} and \ref{fig:posts_stats}. In this case, the SBI$_\mathrm{sbc}$ and SBI$_\mathrm{loss}$ posteriors are systematically broader than the corresponding LBI posteriors, with larger values of $\sigma$ and wider credible intervals. More importantly, the combined SBI constraints are often weaker than those obtained from either $P$ or $B$ alone. This behavior is unexpected for a well-calibrated joint analysis and indicates that the SBI posterior estimator is underconfident in this regime.

\subsubsection{Theoretical expectations for data combination}
To interpret this behavior, we briefly review the theoretical expectations for combining datasets. From an information-theoretic perspective, the information gained about a parameter $\theta$ from data $x$ is defined as the reduction of entropy of the posterior distribution relative to the prior,
\begin{align}
    \mathcal{I}(\theta;x) &= H(\theta) - H(\theta|x),
\end{align}
where the entropy of the posterior for a given data realization is
\begin{align}
    H(\theta|x) &= -\int p(\theta|x)\,\log p(\theta|x)\,\mathrm{d}\theta.
\end{align}
The prior and posterior entropies quantify the initial and remaining uncertainties, so that the information gained corresponds to the uncertainty removed by the data. This quantity is equivalent to the KLD between posterior and prior,
\begin{align}
    \mathcal{I}(\theta;x) &= D_\mathrm{KL}\!\left[p(\theta|x)\,||\,p(\theta)\right] \geq 0.
\end{align}
For a flat prior $\mathcal{U}(a,b)$, one has $H(\theta)=\log(b-a)$, while for a Gaussian posterior $\mathcal{N}(\mu,\sigma^2)$, $H(\theta|x)=\tfrac{1}{2}\log(2\pi e \sigma^2)$. In general, however, the entropy depends on the full shape of the posterior distribution.

The mutual information is obtained by averaging the information gain over the data-generating process,
\begin{equation}
    I(\theta;x)
    =
    \mathbb{E}_x\!\left[\mathcal{I}(\theta;x)\right]
    =
    H(\theta)
    -
    \int p(x)\, H(\theta|x)\,\mathrm{d}x,
\end{equation}
where $p(x)$ denotes the prior predictive distribution. A classic result from information theory is that, for two datasets $A$ and $B$,
\begin{align}
    I(\theta;A,B) \ge \max\!\left[I(\theta;A), I(\theta;B)\right],
    \label{eq:infogrows}
\end{align}
which follows from the chain rule and the non-negativity of conditional mutual information. This property holds for mutual information, but does not apply to the realization-dependent quantity $\mathcal{I}(\theta;x)$, which can fluctuate across datasets and may decrease when combining data in the presence of statistical fluctuations or tension between constraints.

If the expectation is instead taken over data realizations generated at a fixed parameter value $\theta_0$, i.e., $x \sim p(x|\theta_0)$, one obtains the expected information gain at fixed truth,
\begin{equation}
    \tilde{I}(\theta_0)
    =
    \mathbb{E}_{x|\theta_0}\!\left[\mathcal{I}(\theta;x)\right].
\end{equation}
This quantity can be interpreted as an average over a slice of the prior predictive distribution. Unlike mutual information, it does not in general satisfy exact monotonicity relations when combining datasets.

A relation similar to Eq.~(\ref{eq:infogrows}) is, however, expected to hold asymptotically. Under regularity conditions and in the regime where the Bernstein--von Mises theorem applies, the posterior distribution at fixed true parameter value $\theta_0$ approaches a Gaussian with covariance given by the inverse Fisher information. In this limit, if the combined dataset carries more Fisher information than either dataset individually, the posterior variance decreases accordingly. One therefore expects
\begin{equation}
\sigma(\theta|A,B)
\leq
\min\!\left[\sigma(\theta|A),\sigma(\theta|B)\right],
\end{equation}
with deviations in finite-sample or non-Gaussian settings.

\subsubsection{Empirical evidence for loss of constraining power}

We now quantify how strongly these asymptotic expectations are violated in practice. In particular, we measure how often the joint $P+B$ constraint improves upon the individual ones. For the LBI posteriors, $84.3\%$ of realizations satisfy $\sigma_{P+B}\leq\min(\sigma_P,\sigma_B)$, and $86.0\%$ satisfy $C_{68,P+B}\leq\min(C_{68,P},C_{68,B})$. For the SBI$_\mathrm{sbc}$ posteriors, these fractions drop to $6.7\%$ and $6.5\%$, while for SBI$_\mathrm{loss}$ the fractions slightly improve at $19.8\%$ and $16.0\%$.

This stark discrepancy indicates that the SBI model for $P+B$ is systematically underconfident, failing to recover the expected gain in constraining power when combining datasets. Crucially, this behavior is not captured by coverage-based diagnostics, demonstrating that correct coverage does not guarantee reliable posterior calibration, nor an accurate description of posterior tails or joint information content.

This degradation does not appear to be a simple consequence of data-vector dimensionality or limited training data. The $P+B+\mathrm{WST}$ data vector is even higher-dimensional, yet does not exhibit the same behavior, despite showing comparable performance under coverage-based diagnostics. For comparison, LBI yields
$\sigma_{P+B+\mathrm{WST}}\leq\min(\sigma_P,\sigma_B,\sigma_{\rm WST})$
for $89.3\%$ of the realizations; the same fraction also satisfies
$C_{68,P+B+\mathrm{WST}}\leq\min(C_{68,P},C_{68,B},C_{68,\mathrm{WST}})$.
For SBI, these fractions decrease to $48.0\%$ and $49.3\%$,
respectively, representing a substantial improvement over the $P+B$
case. This suggests that the issue is not driven by dimensionality, but
instead reflects a more specific limitation of the validation metrics:
they do not reliably detect this particular failure mode of the
posterior.

\begin{figure}
    \centering
    \includegraphics[width=\linewidth]{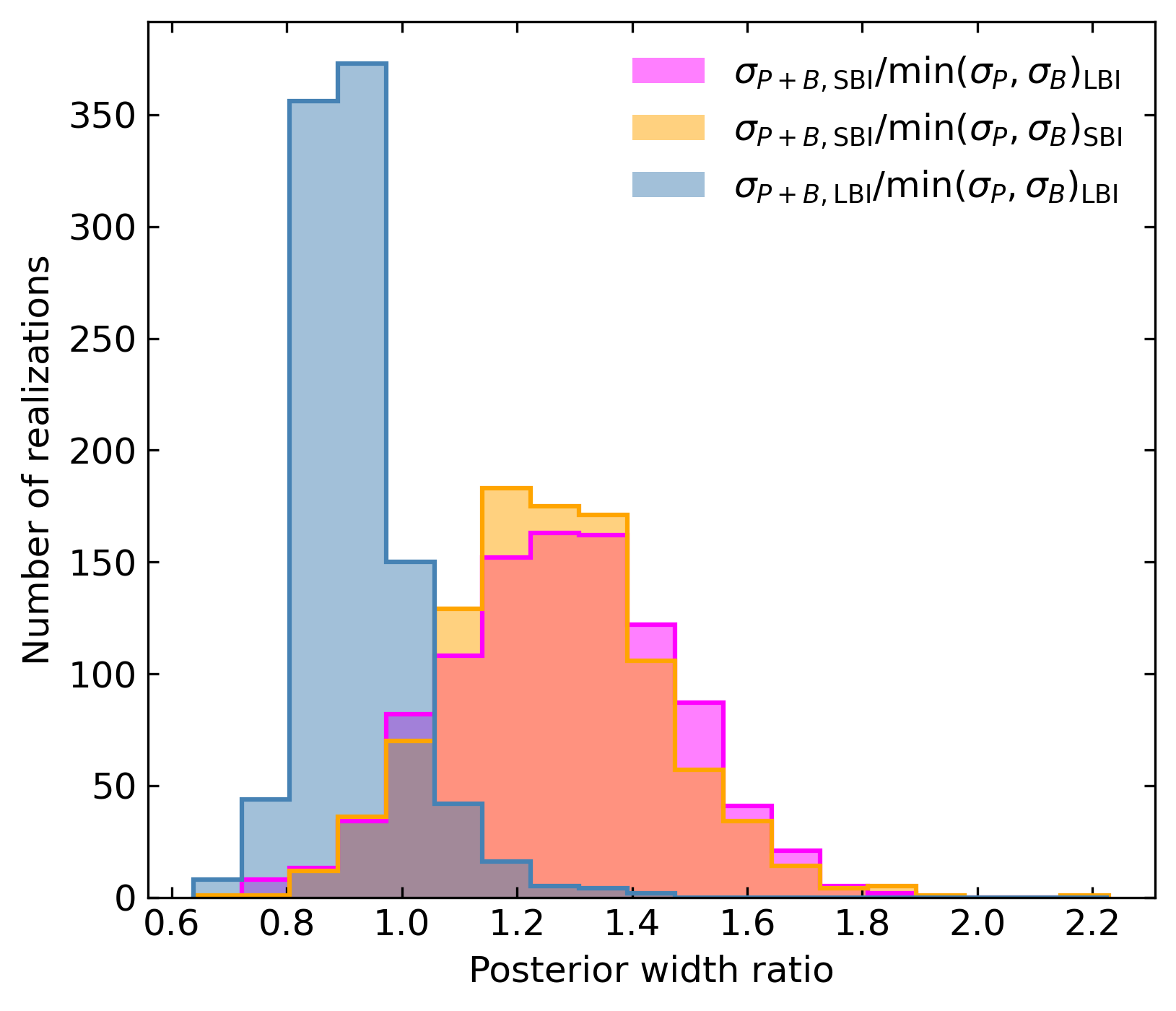}
    \caption{Distributions of the ratio
$\sigma_{P+B}/\min(\sigma_P,\sigma_B)$, where
$\min(\sigma_P,\sigma_B)$ denotes the tighter of the individual
constraints from $P$ and $B$. The legend
identifies whether the numerator and denominator are obtained from LBI
or SBI. The mixed LBI--SBI comparison tests whether differences in
posterior widths between the two inference methods can account for the
apparent underconfidence of the SBI $P+B$ posteriors.}
    \label{fig:ratios_dist}
\end{figure}

One possible explanation for the low SBI pass rate is that the criterion $\sigma_{P+B}\leq\min(\sigma_P,\sigma_B)$ is intrinsically more
difficult to satisfy if the individual constraints from $P$ and $B$ are
already tighter for SBI than for LBI. As suggested by the anonymous referee,
this could occur if SBI is able to exploit features of the individual sampling distributions of $P$ and $B$ that are not captured by the Gaussian likelihood adopted in LBI. To test whether this effect can account for
the observed discrepancy, we compare the SBI constraint on $P+B$ with
the corresponding constraints obtained from $P$ and $B$ individually
using LBI. Figure~\ref{fig:ratios_dist} shows the distribution of the
resulting ratios, together with the corresponding distributions obtained
when both the joint and individual constraints are derived entirely from
SBI or entirely from LBI. Although the mixed LBI--SBI comparison is
somewhat broader than the pure SBI case, the difference is too small to
explain the observed behavior of the $P+B$ posteriors.
We therefore conclude
that the smaller fraction of SBI realizations satisfying 
$\sigma_{P+B}\leq\min(\sigma_P,\sigma_B)$
cannot be attributed simply to tighter constraints from $P$ and $B$
individually, but instead reflects genuine underconfidence of the SBI
$P+B$ posteriors.

\subsection{The information content of large-scale structure}

The question of whether cosmological information extends beyond standard low-order summary statistics, such as the power spectrum and bispectrum, has been the subject of ongoing debate, primarily in the context of parameters like the amplitude of clustering, $A_\mathrm{s}$ or $\sigma_8$. While field-level analyses have been argued to extract additional information beyond low-order statistics \citep{NguSchTuc2411}, effective field theory–based approaches often suggest that most of the accessible information is already captured by $P+B$ on sufficiently large scales, where a perturbative treatment is valid, once modeling uncertainties are taken into account \citep[e.g.,][and references therein]{AkiSimChe2509}.

In the context of PNG, our results provide empirical support for the former perspective. We find that including WST coefficients substantially improves the constraints on $f_\mathrm{NL}$, in agreement with the Fisher forecasts of \cite{Peron_2024}, indicating that a non-negligible fraction of the available information is not captured by $P+B$ alone. Importantly, this conclusion holds even though our analysis is restricted to modes with $k_\mathrm{max} = 0.141\,h\,\mathrm{Mpc}^{-1}$ for all summary statistics. This implies that the additional constraining power provided by the WST coefficients arises from large-scale modes within the perturbative regime.

\section{Conclusions}
\label{sec:conclusions}

Simulation-based inference has emerged as a powerful alternative to standard likelihood-based methods in cosmology, particularly in settings where analytical likelihoods are unavailable or difficult to model accurately. Its flexibility makes it especially attractive for analyses based on complex observables and nonlinear summary statistics, such as WST coefficients or field-level neural summaries. At the same time, the reliability of SBI posterior estimates remains an open question. In practice, available diagnostics primarily focus on coverage, i.e.,\ whether the true parameter value falls within posterior credible regions with the expected frequency across many realizations. While this provides a useful and well-motivated validation of posterior calibration under the prior predictive distribution, it does not guarantee correct behavior at a fixed parameter value. In particular, models that pass coverage tests can still exhibit systematic biases or the misestimation of uncertainties under repeated sampling at a fixed truth,
motivating complementary frequentist validation.

In this work, we used the inference of a single parameter, $f_\mathrm{NL}$, from the power spectrum, bispectrum, and WST coefficients of the dark matter halo field as a controlled test case. We compared posterior estimates obtained from standard LBI (assuming a Gaussian likelihood) to those obtained from SBI using CNRE, using 1000 fiducial test simulations. Our main findings are as follows:

\begin{enumerate}
\item Among the summary statistics considered here, only the power spectrum $P$ consistently passes all multivariate normality tests, making it the cleanest case in which the Gaussian-likelihood assumption underlying our LBI analysis is well justified. The bispectrum, WST coefficients, and their combinations all show measurable departures from multivariate Gaussianity 
(Table~\ref{tab:mnt_mvn_results}).
\item Despite this, posterior estimates obtained from SBI and LBI show strong agreement in the first moment: posterior means are highly correlated across realizations and consistent, on average, with the true value (Figs.~\ref{fig:big_plot_main} and~\ref{fig:posts_stats}). The standard deviation is also broadly consistent on average, although its realization-by-realization agreement is weaker and degrades for more complex data vectors. This indicates that both methods capture the central location and overall scale of the posterior similarly, even in the presence of moderate non-Gaussianity, but they respond differently to individual noise realizations.
\item The agreement partially extends to higher-order structures. The skewness remains strongly correlated between SBI and LBI (except for WST), indicating a similar description of posterior asymmetry. In contrast, the kurtosis shows substantial discrepancies, with little to no correlation between the two methods and even differences in sign (Fig.~1 and Fig.~\ref{fig:posts_stats}, bottom panels). This highlights significant differences in how the tails of the posterior are modeled, with SBI posteriors often appearing more weakly constrained or heavy-tailed in a realization-dependent manner.
\item The clearest example is the combined $P+B$ data vector. In this case, SBI posteriors are systematically underconfident relative to LBI and can yield weaker constraints than either $P$ or $B$ alone. This behavior is not detected by the standard calibration diagnostics used in this work, showing that passing coverage-based tests does not guarantee that the inferred posterior shape is reliable for frequentist validation. These effects can lead to sizable discrepancies in individual posterior estimates (Fig.~\ref{fig:p_posts_comp}).
\item The WST coefficients provide substantially stronger constraints on $f_\mathrm{NL}$ than the $P+B$ combination alone. This agrees qualitatively with the literature, and the gain persists even when the analysis is restricted to very large scales, $k_\mathrm{max}=0.141\,h\,\mathrm{Mpc}^{-1}$.

\end{enumerate}

Our results indicate that current validation pipelines are not sufficient to guarantee reliable posterior shapes. More broadly, they highlight two key points. First, higher-order summary statistics such as the WST contain significant additional information on PNG beyond traditional two- and three-point functions, even on large scales. Second, the increasing use of such summaries makes robust posterior validation an essential component of cosmological SBI pipelines. In particular, diagnostics that probe the behavior of posteriors at a fixed parameter value and that are sensitive to their full shape, beyond coverage alone, will be necessary for reliable uncertainty quantification. Developing such tools---for example through improved tail-sensitive diagnostics, likelihood-free goodness-of-fit tests, or hybrid validation strategies combining Bayesian and frequentist perspectives---will be key for SBI to fully realize its potential for precision cosmology.

\begin{acknowledgements}
We thank the anonymous referee for a thorough reading of the manuscript and for helpful suggestions that improved it. This work was funded by the Deutsche Forschungsgemeinschaft (DFG, German Research Foundation) under Germany’s Excellence Strategy EXC 3037 - 533607693. We acknowledge access to the Marvin HPC cluster at the University of Bonn. We thank the \textsc{Quijote} and \textsc{Quijote-PNG} teams for making their simulation data publicly available. This work made use of the following open-source software: \textsc{Pylians}, \textsc{Kymatio}, \textsc{PySpectrum}, \textsc{sbi}, \textsc{Optuna}, and the \textsc{mnt} library in R. TA is a member of the International Max Planck Research School (IMPRS) for Astronomy and Astrophysics and the Bonn–Cologne Graduate School (BCGS) of Physics and Astronomy.
\end{acknowledgements}

\bibliographystyle{bibtex/aa}
\bibliography{main}

\begin{appendix}

\section{WST coefficients in Fourier space}
\label{app:wst_ps}

\subsection{Fourier representation of solid harmonic wavelets}

The Fourier transform of the solid harmonic wavelets can be written as
\begin{align}
\widetilde{\psi}_{j \ell m}(\bm{k})
&= \int \psi_{j\ell m}(\bm{r})\, \mathrm{e}^{i \bm{k}\cdot\bm{r}} \mathrm{d}^3r
=4\pi \,i^{\ell}\, Y_{\ell m}(\hat{\bm{k}})\,I_{j\ell}(k),
\end{align}
where the radial component is
\begin{align}
I_{j\ell}(k)&=\frac{1}{(2\pi)^{3/2}\,\sigma_j^3}\, \int_0^\infty \mathrm{e}^{-r^2/2\sigma_j^2} \,r^{\ell+2} \,j_{\ell}(kr) \, \mathrm{d}r \nonumber \\
&=\frac{\sigma_j^\ell}{(2\pi)^{3/2}}\, \int_0^\infty \mathrm{e}^{-w^2/2} \,w^{\ell+2} \,j_{\ell}(k\sigma_j w) \, \mathrm{d}w.
\label{eq:FTI}
\end{align}
Using the relation between spherical and cylindrical Bessel functions,
\begin{align}
j_\ell(z) = \sqrt{\frac{\pi}{2z}}\, J_{\ell + \frac{1}{2}}(z),
\end{align}
the radial integral can be evaluated analytically \citep[see integral 6.631.4 in][]{gradshteyn2007}, yielding
\begin{align}
I_{j\ell}(k)
=\frac{\sigma_j^{\ell}}{4\pi}\, (k\sigma_j)^{\ell}\,\mathrm{e}^{-(k\sigma_j)^2/2}.
\end{align}
The Fourier transform therefore takes the separable form
\begin{align}
\widetilde{\psi}_{j\ell m}(\bm{k}) =
(i \sigma_j)^{\ell}\, (k\sigma_j)^{\ell}\, \mathrm{e}^{-(k\sigma_j)^2/2} \,Y_{\ell m}(\hat{\bm{k}})
\equiv \widetilde{W}_{j\ell}(k)\, Y_{\ell m}(\hat{\bm{k}}),
\label{eq:FTwavelet}
\end{align}
which makes explicit the factorization into radial and angular components.

\subsection{Connection between low-order WST coefficients and the power spectrum}

For $q=2$, the zeroth-order coefficient reduces to the variance of the overdensity field,
\begin{align}
    S^{(0)}=\left\langle \delta^2(\bm{x}) \right\rangle
    =\frac{1}{2\pi^2}\int_0^\infty k^2 P(k) \,\mathrm{d}k,
    \label{eq:s0power}
\end{align}
showing that $S^{(0)}$ corresponds to the total integrated power.

Using the convolution theorem, the definition of the power spectrum, and the orthonormality of spherical harmonics, the first-order coefficients can be written as
\begin{align}
S^{(1)}_{j \ell} =
\frac{(2\ell+1)}{(2\pi)^3}
\int_0^\infty k^2 P(k)\,
\left|\widetilde{W}_{j \ell}(k)\right|^2\,\mathrm{d}k.
\label{eq:s1power}
\end{align}
This expression shows that $S^{(1)}_{j\ell}$ corresponds to a wavelet-weighted (i.e., band-pass filtered) version of the power spectrum. In this sense, the first-order scattering coefficients provide a decomposition of the variance across both spatial scales and angular frequencies.

\subsection{Shot-noise contributions to low-order WST coefficients}
\label{sec:WST_shot_noise}

For discrete tracers, shot noise introduces an additive contribution to the power spectrum,
\begin{align}
    P_\mathrm{t}(k)=P(k)+\frac{1}{\bar n}.
\end{align}
From Eq.~(\ref{eq:s0power}), the corresponding contribution to $S^{(0)}$ is formally divergent, reflecting the ultraviolet divergence of white noise. In practice, this divergence is regulated by the finite resolution of the density field.

For the first-order coefficients, the shot-noise contribution remains finite,
\begin{align}
   S^{(1\mathrm{t})}_{j\ell}= S^{(1)}_{j\ell}+\frac{\Phi_{j\ell}}{\bar{n}},
\end{align}
with
\begin{align}
    \Phi_{j\ell}&= \frac{(2\ell+1)}{(2\pi)^3}
    \int_0^\infty k^2\,
    \left|\widetilde{W}_{j \ell}(k)\right|^2\,\mathrm{d}k\nonumber \\
    &=\frac{(2\ell+1)}{(2\pi)^3}\,\sigma_{j}^{-(2\ell+3)}\,
    \frac{\Gamma(\ell + \tfrac{3}{2})}{2}.
\end{align}

\subsection{Second-order coefficients and connection to higher-order statistics}
\label{app:S2_exp}
To elucidate the connection between second-order scattering coefficients and higher-order statistics, we consider the first-layer rotationally invariant field
\begin{align}
\rho_{j_1\ell}(\bm{x}) \equiv \left[\sum_{m=-\ell}^{\ell} \left| \delta * \psi_{j_1 \ell m}(\bm{x}) \right|^2 \right]^{1/2}.
\end{align}
We define
\begin{align}
A_{j_1\ell} \equiv \left\langle \sum_{m=-\ell}^{\ell} \left| \delta * \psi_{j_1 \ell m} \right|^2 \right\rangle = \left\langle \rho_{j_1\ell}^2 \right\rangle,
\end{align}
and the fluctuation field
\begin{align}
\epsilon_{j_1\ell}(\bm{x}) \equiv \sum_{m=-\ell}^{\ell} \left| \delta * \psi_{j_1 \ell m}(\bm{x}) \right|^2 - A_{j_1\ell}.
\end{align}
Expanding
\begin{align}
\rho_{j_1\ell}(\bm{x}) = \left[ A_{j_1\ell} + \epsilon_{j_1\ell}(\bm{x}) \right]^{1/2}
\end{align}
around $A_{j_1\ell}$ yields
\begin{align}
\rho_{j_1\ell}(\bm{x})
=
\sqrt{A_{j_1\ell}}
+
\frac{\epsilon_{j_1\ell}(\bm{x})}{2\sqrt{A_{j_1\ell}}}
-
\frac{\epsilon_{j_1\ell}^2(\bm{x})}{8A_{j_1\ell}^{3/2}}
+ \cdots.
\end{align}
Since the wavelets have zero mean, the constant term does not contribute upon convolution. The second-order coefficient
\begin{align}
S^{(2)}_{j_1 j_2 \ell}
=
\left\langle
\sum_{m=-\ell}^{\ell}
\left|
\rho_{j_1\ell} * \psi_{j_2 \ell m}
\right|^2
\right\rangle
\end{align}
is therefore, at leading order,
\begin{align}
S^{(2)}_{j_1 j_2 \ell}
\approx
\frac{1}{4A_{j_1\ell}}
\left\langle
\sum_{m=-\ell}^{\ell}
\left|
\epsilon_{j_1\ell} * \psi_{j_2 \ell m}
\right|^2
\right\rangle.
\end{align}
\begin{figure*}
    \includegraphics[width=17cm]{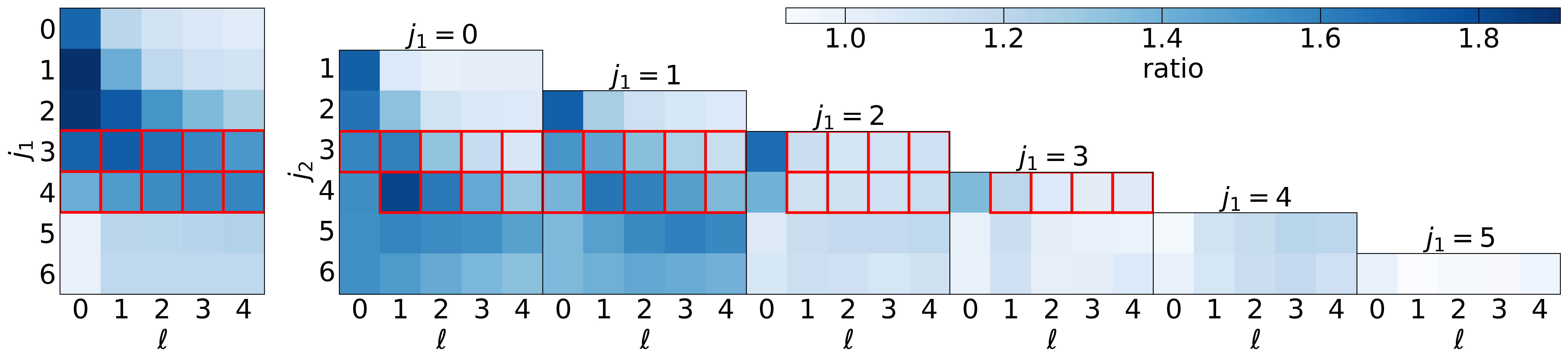}
    \caption{Ratio of the standard deviations of the WST coefficients measured from simulations with varying $f_\mathrm{NL}$ to those from simulations with fixed $f_\mathrm{NL}=0$ (left: first-order; right: second-order). Larger ratios indicate greater sensitivity to $f_\mathrm{NL}$. The coefficients retained in our analysis are highlighted in red.}
    \label{fig:s1s2_sigs}
\end{figure*}
The field $\epsilon_{j_1\ell}$ is quadratic in $\delta$, so the leading contribution to $S^{(2)}$ is quartic in the density field. In Fourier space,
\begin{align}
S^{(2)}_{j_1 j_2 \ell}
\approx
\frac{1}{4A_{j_1\ell}}
\int 
\sum_{m=-\ell}^{\ell}
\left| \tilde{\psi}_{j_2 \ell m}(\bm{q}) \right|^2
\, P_{\epsilon_{j_1\ell}}(\bm{q})\,
\frac{\mathrm{d}^3 q}{(2\pi)^3}
\;,
\end{align}
where $P_{\epsilon_{j_1\ell}}$ is the power spectrum of the quadratic field $\epsilon_{j_1\ell}$. This decomposes into a connected contribution involving the trispectrum of $\delta$ and a disconnected Gaussian term proportional to $P \times P$,
\begin{align}
P_{\epsilon_{j_1\ell}} = P^{(T)}_{\epsilon_{j_1\ell}} + P^{(G)}_{\epsilon_{j_1\ell}}.
\end{align}
Thus, second-order scattering coefficients probe fluctuations of the local wavelet power and are sensitive to mode coupling encoded in higher-order correlation functions. At leading order, they correspond to wavelet-weighted integrals of the trispectrum and of disconnected $P \times P$ terms. Higher-order terms in the expansion generate additional contributions involving higher-order polyspectra. In particular, all even-order correlation functions contribute, including terms associated with the 6-point, 8-point, and higher-order polyspectra, reflecting the intrinsically nonlinear nature of the scattering transform. This mixing of contributions implies that scattering coefficients provide a compressed representation of higher-order statistics, capturing information that would otherwise require the explicit measurement of a large number of polyspectra.

\subsection{Selection of WST coefficients}
\label{appndx:wst}

The full set of WST coefficients is highly redundant, allowing for a significant reduction in the dimensionality of the data vector without loss of information. To identify the most informative coefficients for constraining $f_\mathrm{NL}$, we compare their standard deviation across simulations with varying $f_\mathrm{NL}$ to that of fiducial simulations with $f_\mathrm{NL}=0$. Figure~\ref{fig:s1s2_sigs} shows that coefficients at very small and large wavelet scales contribute little additional information to that within the remaining coefficients. Restricting to $3 \le j_1 \le 4$ (and applying the same range to $j_2$ for second-order coefficients) leaves the inference unchanged while significantly reducing dimensionality. This is consistent with the fact that local PNG primarily induces a scale-dependent bias on large scales, while our sharp-$k$ filters also remove modes outside of $k \in [0.009, 0.141]\,h\,\mathrm{Mpc}^{-1}$.

\begin{figure}
    \centering
    \resizebox{\hsize}{!}{\includegraphics{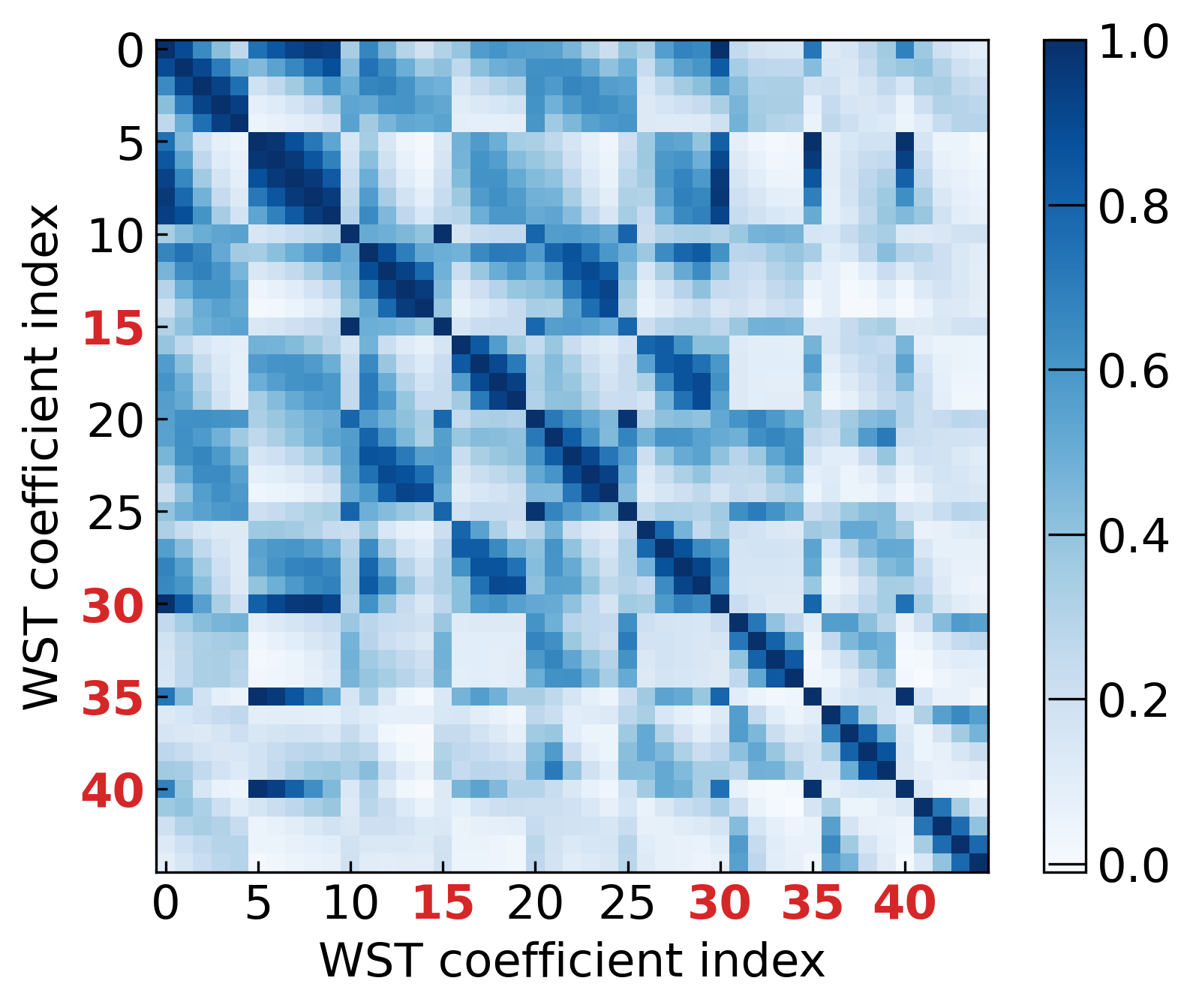}}
    \caption{Correlation matrix of the WST coefficients after restricting to $3 \le j_1 \le 4$ (first order) and $3 \le j_2 \le 4$ (second order). The ordering follows Fig.~A1, with first-order coefficients appearing first (indices 0–9), followed by second-order coefficients. Indices are grouped such that sets of five correspond to different $\ell$ at fixed scales $j_{1,2}$, with second-order coefficients ordered by $j_2$, $j_1$, and $\ell$. Highly correlated coefficients (indices 15, 30, 35, and 40; shown in red) are excluded from the analysis.}
    \label{fig:corr_mats}
\end{figure}

We further remove strongly correlated coefficients, identified from the correlation matrix, in particular first- and second-order coefficients with $\ell=0$ and matching scales ($j_1^a = j_2^b$), as well as additional redundant configurations. Figure~\ref{fig:corr_mats} shows that such coefficients are highly correlated, which necessitates excluding most of them to get a well-conditioned correlation matrix for the remaining set of 41 WST coefficients, highlighted in Fig.~\ref{fig:s1s2_sigs}.

\section{SBI optimization and diagnostics}
\label{appndx:diagnostics}

\subsection{Optimizing the architecture and training procedure}

Our SBI pipeline uses the default CNRE implementation provided by the
\textsc{sbi} package \citep{cnre_paperb}. The classifier is a fully
connected Multi-Layer Perceptron with three hidden layers of 128 neurons
each and ReLU activation functions. It takes the concatenated parameter
vector $\theta$ and data vector $x$ as input and produces the logits
required by the CNRE contrastive loss.

To obtain reliable SBI posterior estimators, we perform hyperparameter optimization using the \textsc{optuna} package \citep{optuna}. The search space includes the learning rate $\eta \in [5 \times 10^{-4}, 10^{-2}]$, batch size $B \in [4, 128]$, number of contrastive samples $K \in [2,20]$, and the minimum number of training epochs before early stopping $n_\mathrm{ep} \in [10,50]$. We train a separate SBI model for each data vector and optimize its hyperparameters independently.

\begin{figure}
    \centering
    \resizebox{\hsize}{!}{\includegraphics{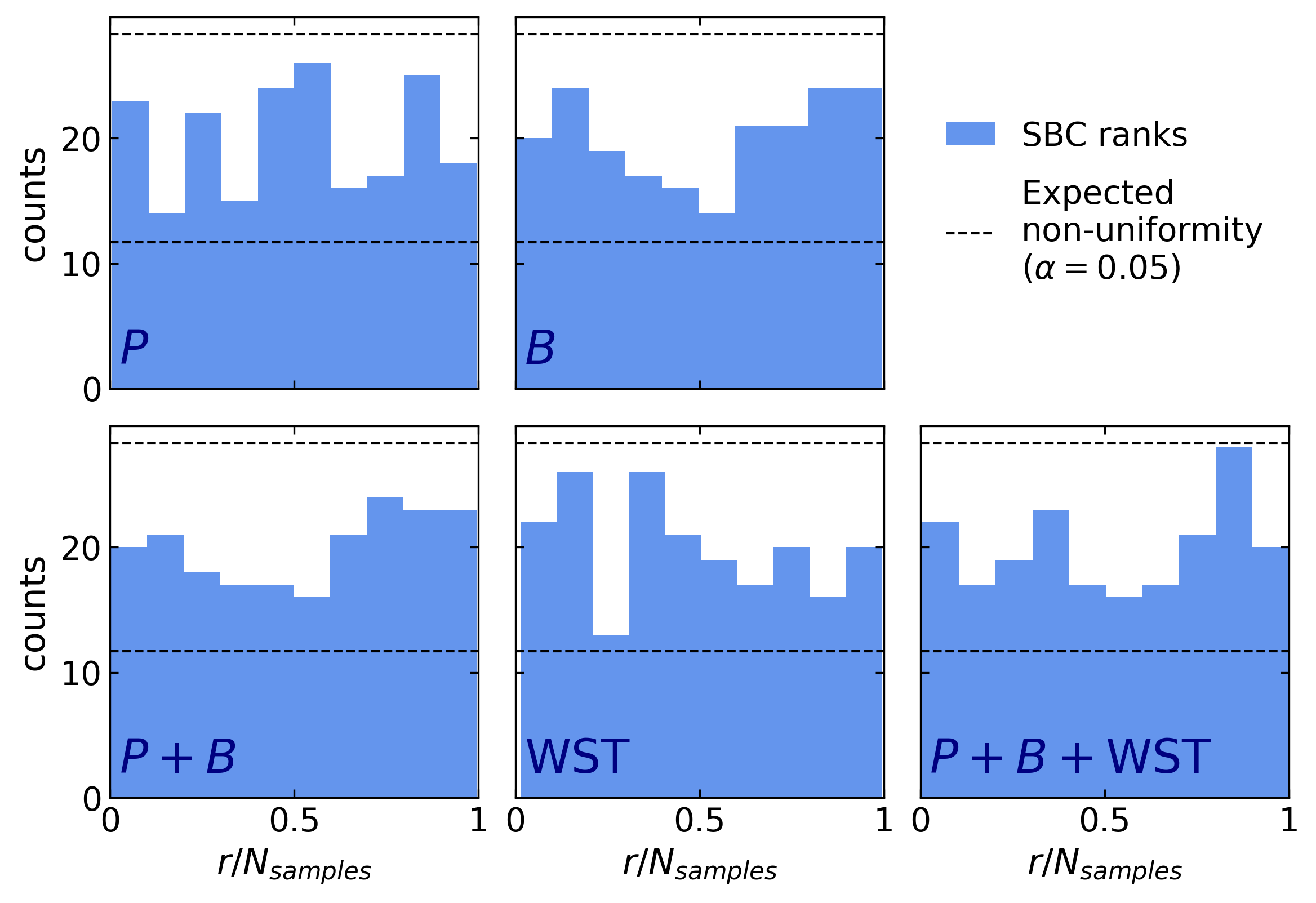}}
    \caption{SBC rank histograms for the $\mathrm{SBI}_\mathrm{sbc}$ models, evaluated on 200 validation realizations. The dashed lines indicate the $95\%$ confidence envelope ($\alpha=0.05$) for a uniform distribution. Ranks $r$ are normalized by the number of posterior samples per realization, $N_\mathrm{samples}$.}
    \label{fig:sbc_figs}
\end{figure}

We consider two optimization strategies. First, we directly target posterior calibration by minimizing deviations of the SBC rank distribution from uniformity. Specifically, we minimize the maximum deviation of the SBC rank empirical CDF from the diagonal, normalized by the expected $95\%$ confidence envelope. We perform 300 optimization trials and select the configuration that minimizes this objective, yielding the $\mathrm{SBI}_\mathrm{sbc}$ models. Figure~\ref{fig:sbc_figs} shows the corresponding SBC rank histograms for 200 test realizations, consistent with uniformity within expected statistical fluctuations.

\begin{figure}
    \centering
    \resizebox{\hsize}{!}{\includegraphics{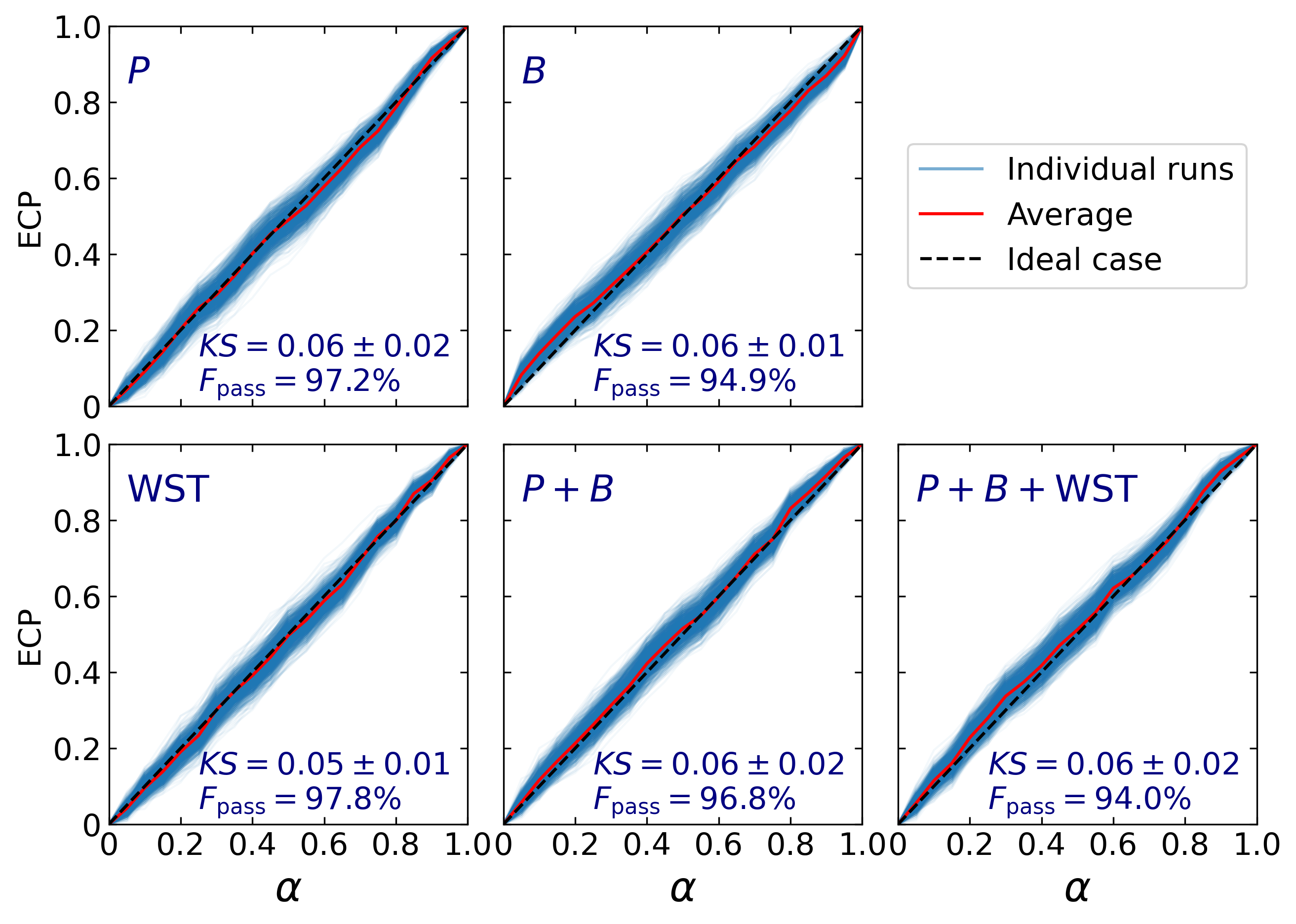}}
    \caption{TARP calibration results for the $\mathrm{SBI}_\mathrm{sbc}$ models across different summary statistics. Blue lines show individual runs, and the red line their mean. Reported values include the median KS statistic, its $68\%$ interval, and the fraction of runs consistent with uniform coverage at $95\%$ confidence.}
    \label{fig:tarp_figs}
\end{figure}

We further validate these models using the TARP test, shown in Fig.~\ref{fig:tarp_figs}. For each summary statistic, we perform $N=1000$ independent runs, drawing reference points from the prior and evaluating coverage over the 200 test realizations. Calibration is assessed using a Kolmogorov--Smirnov (KS) test against uniformity. We report the median KS statistic, the width of its central $68\%$ interval, and the fraction of runs passing the KS test at the $95\%$ level, $F_\mathrm{pass}$. All summary statistics yield median KS values in the range $0.05$--$0.07$ and pass rates between $95\%$ and $98\%$, indicating well-calibrated $\mathrm{SBI}_\mathrm{sbc}$ posteriors.

As a comparison, we also optimize using the standard cross-entropy validation loss, yielding $\mathrm{SBI}_\mathrm{loss}$ models. Although these do not explicitly target calibration, they also pass SBC and TARP tests and produce robust posterior estimates. Differences between the two approaches are discussed in Section~\ref{subsec:comparison}.

\subsection{Convergence with training-set size}

\cite{Bairagi_2025} showed that the number of
simulations required to train a converged SBI model depends on both the
dimensionality of the data vector and the dimensionality of the
parameter space. For a power spectrum with $n_\xi=128$ data points and
a five-dimensional parameter space, they found that at least
$n_{\rm train}\sim3000$ simulations are required to obtain converged
posterior estimates. In our case, the data vectors are considerably
smaller ($n_\xi\in\{7,41,62,69,110\}$) and the inference problem is
one-dimensional. We therefore expect convergence to be achieved with
fewer training simulations.

\begin{figure}
    \centering
    \includegraphics[width=\linewidth]{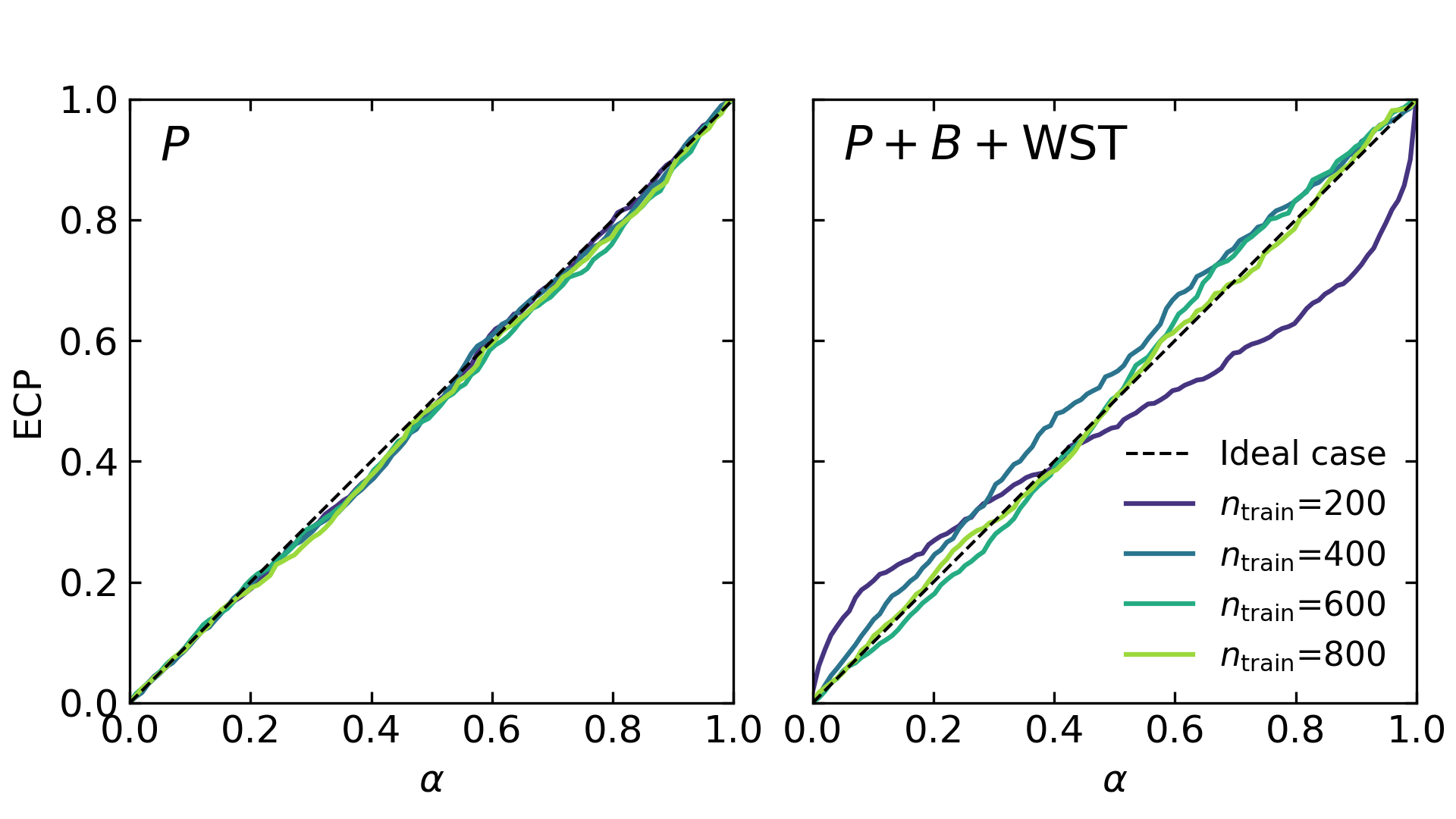}
    \caption{ TARP coverage for SBI models trained on the $P$ (left) and
$P+B+\mathrm{WST}$ (right) data vectors using different numbers of
training simulations. The curves show the mean coverage from 1000 TARP
realizations.}
    \label{fig:tarp_convergence}
\end{figure}

To test this expectation, we use our shortest data vector, $P$, as a
reference case and our longest one, $P+B+\mathrm{WST}$, as a benchmark
for the most demanding scenario. Figure~\ref{fig:tarp_convergence}
shows the TARP coverage obtained for models trained with
$n_{\rm train}\in\{200,400,600,800\}$ simulations. For $P$, all models
pass the TARP coverage test, whereas for $P+B+\mathrm{WST}$ this is
only the case for $n_{\rm train}\ge600$.

\begin{figure}
    \centering
    \includegraphics[width=0.9\linewidth]{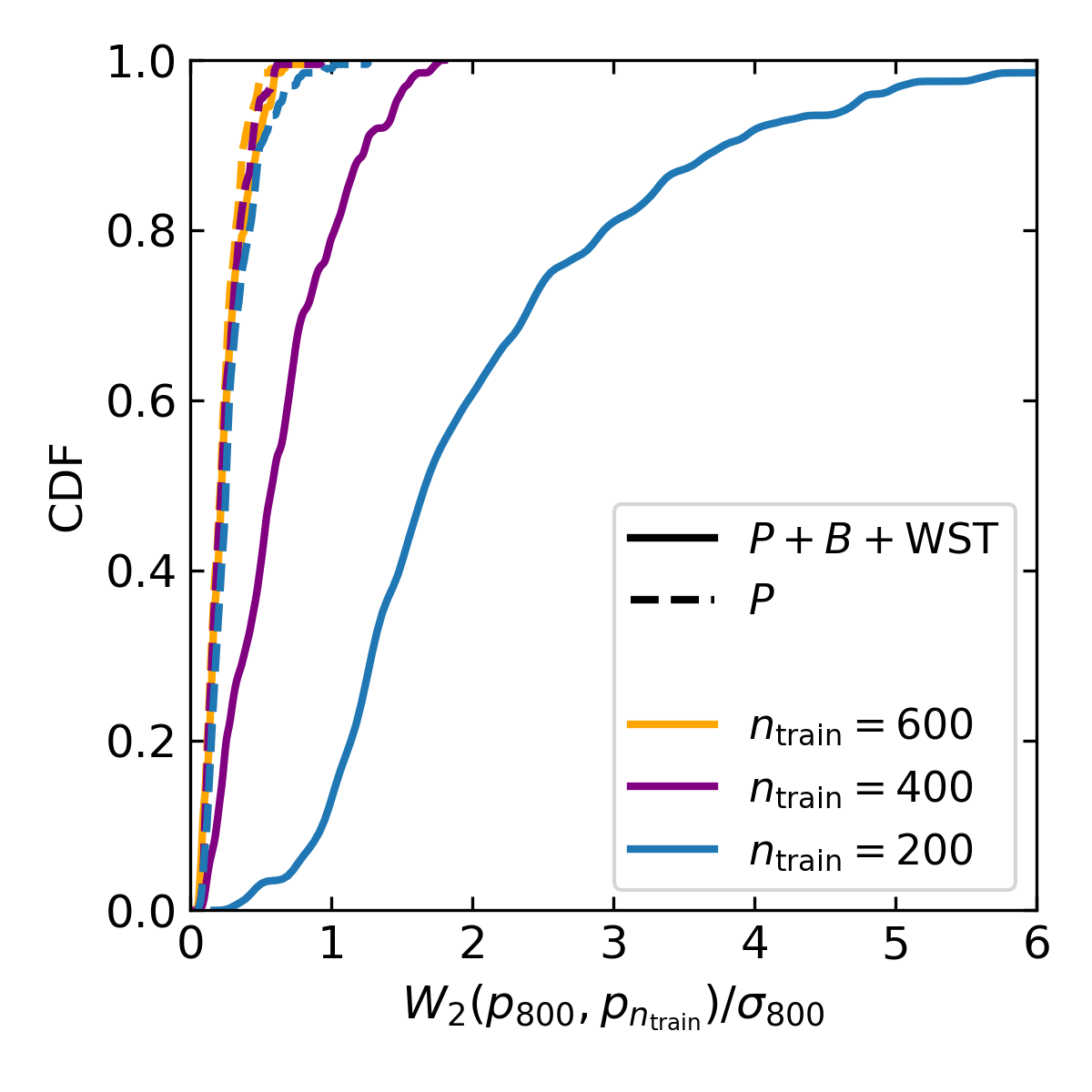}
    \caption{Empirical cumulative distribution functions of the normalized
2-Wasserstein distances between the posterior distributions obtained
with the reference SBI model trained on
$n_{\rm train}=800$ simulations and models trained on smaller training
sets. The $P$ data vector exhibits nearly identical distributions for
all values of $n_{\rm train}$ and is therefore used as a reference for
assessing the convergence of the $P+B+\mathrm{WST}$ model.}
    \label{fig:w2_convergence}
\end{figure}

As a complementary diagnostic, Figure~\ref{fig:w2_convergence} shows the cumulative distribution functions of the normalized 2-Wasserstein distances between the posterior distributions obtained with models trained on
$n_{\rm train}\in\{200,400,600\}$ simulations and those obtained with
our reference model trained on $n_{\rm train}=800$ simulations. For the
$P$ data vector, all training-set sizes produce nearly identical
distributions, whereas for $P+B+\mathrm{WST}$ only the
$n_{\rm train}=600$ model shows comparable agreement with the reference
case. We therefore conclude that, for our longest data vector, the posterior estimates appear to have converged posterior estimates for
$n_{\rm train}\gtrsim600$, making our main analysis based on
$n_{\rm train}=800$ simulations sufficiently robust.

\section{Comparison of SBI architectures}
\label{appndx:inconsistencies}

To assess the robustness of our results with respect to the choice of SBI architecture, we contrast the CNRE method, which we use as our baseline due to its robustness for high-dimensional data vectors, with two alternative approaches: Neural Posterior Estimation \citep[\textsc{NPE\_C},][]{NPE_C} implemented with Masked Autoregressive Flows (MAF),\footnote{By default, univariate inference in \textsc{sbi} is restricted to a Gaussian flow, in which the network learns only the mean $\mu$ and variance $\sigma$ of the posterior. To enable a more flexible MAF model, we introduce an auxiliary dummy parameter $\chi \sim \mathcal{U}[-1000,1000]$, forcing the model into a multivariate setting and allowing it to capture non-Gaussian features in the marginal distribution of $f_\mathrm{NL}$.} and Balanced Neural Ratio Estimation \citep[\textsc{BNRE},][]{bnre_paper}. The latter extends standard NRE with a regularization term that penalizes deviations from a global balance condition, improving the calibration of the learned likelihood-to-evidence ratio.

For this comparison, we focus on the power spectrum, which provides the simplest, lowest-dimensional data vector. All three models pass the SBC and TARP calibration tests and show comparable performance in this respect.

\begin{figure}
    \centering
    \resizebox{\hsize}{!}{\includegraphics{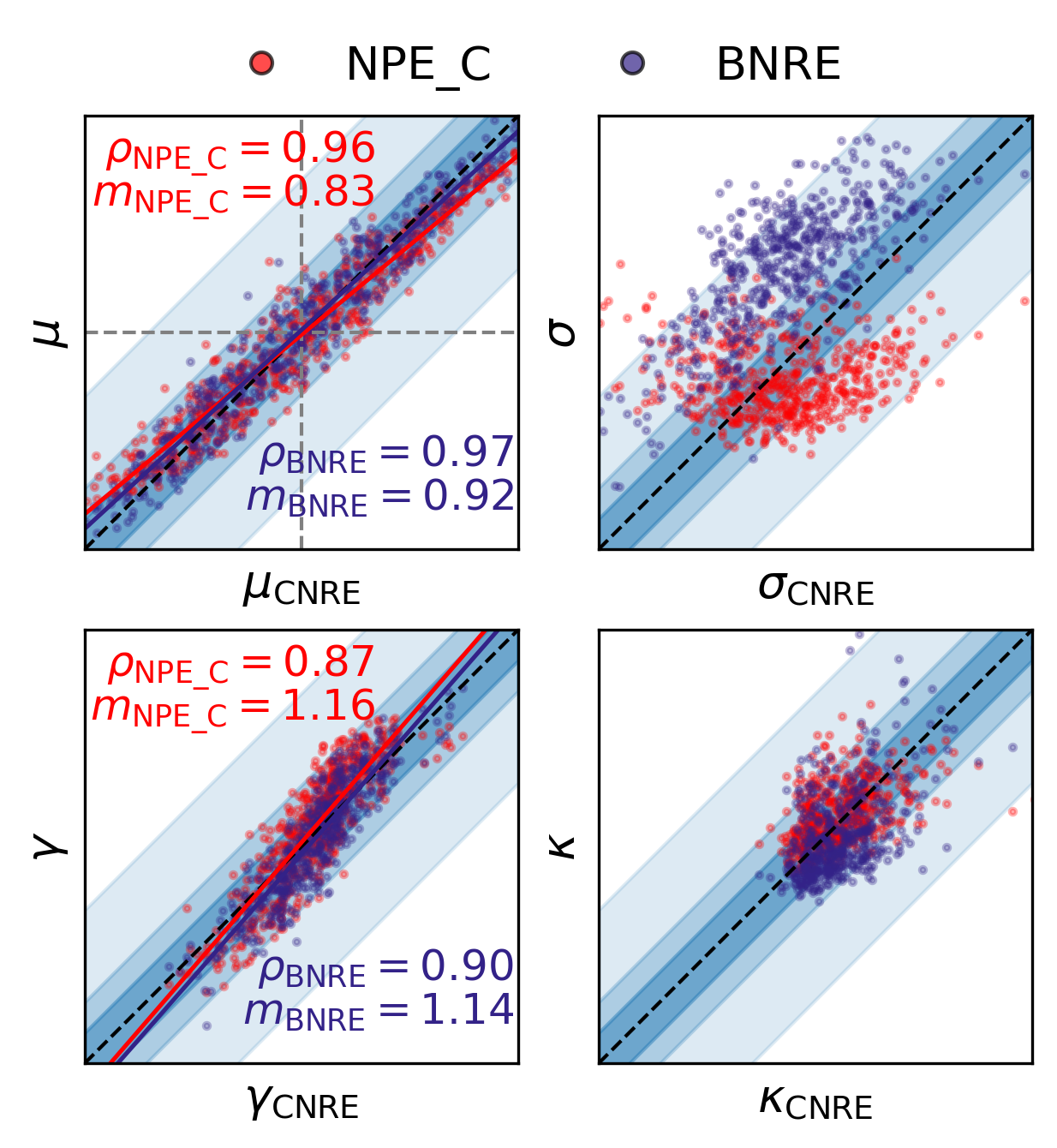}}
    \caption{Comparison of posterior moments for NPE\_C and BNRE relative to CNRE. Same style as Fig.~~\ref{fig:big_plot_main}.}
    \label{fig:sbi_comparison}
\end{figure}

Figure~\ref{fig:sbi_comparison} compares the posterior moments obtained with these methods. The largest discrepancies appear in the standard deviation. While \textsc{BNRE} exhibits mild underconfidence relative to \textsc{CNRE}, \textsc{NPE\_C} tends to produce slightly overconfident posteriors, consistent with previous benchmarks \citep[e.g.,][]{Hermans2021averting}. The deviations typically remain
within $10$--$25\%$.

While such differences may decrease with larger training sets, their presence indicates that, despite passing standard calibration tests, SBI posterior estimates are not uniquely defined in practice.

\end{appendix}

\end{document}